\documentclass[aps,prc,tightenlines,showpacs,preprint,%
amssymb,byrevtex,nofootinbib,superscriptaddress]{revtex4}  

\usepackage{epsfig,bm,dcolumn}

\begin{document}

\title{Quarkonium-Hadron Interactions in Perturbative QCD}
\author{Taesoo Song}
\email[Electronic mail:]{songtsoo@yonsei.ac.kr}
\author{Su Houng Lee}
\email[Electronic mail:]{suhoung@phya.yonsei.ac.kr}
\affiliation{IPAP and Department of Physics,
        Yonsei University, Seoul 120-749 Korea}

\date{\today}

\begin{abstract}
The next to leading order (NLO) quarkonium-hadron cross section is
calculated in perturbative QCD.  The corresponding leading order
(LO) result was performed by Peskin more than 20 years ago using
the operator product expansion (OPE).   In this work, the
calculation is performed using the Bethe-Salpeter amplitude and
the factorization formula.  The soft divergence appearing in the
intermediate stages of the calculations are shown to vanish after
adding all possible crossed terms, while the collinear divergences
are eliminated by mass factorization.  Applying the result to the
Upsilon system, one finds that there are large higher order
correction near the threshold.   The relevance of the present
result to the charmonium case is also discussed.
\end{abstract}


\maketitle

\section{INTRODUCTION}

Twenty five years ago, Peskin \cite{peskin}, and Bhanot and Peskin
\cite{Bhanot} showed that in the heavy quark limit, the
interaction between a hadron and a heavy quark bound state could
be described in perturbative QCD, and calculated the dissociation
cross section of the quarkonium by a hadron to leading order (LO).
According to their calculation, which was based on operator
product expansion (OPE), the  $J/\psi$ dissociation cross section
by a hadron was found to be very small and in the order of
$\mu$barn in the threshold region.   Later the result was
re-derived and the target mass correction
calculated\cite{Kharzeev94,Kharzeev96,Arleo,Oh}, but the
qualitative result remained the same.   Initially, such small
cross section strongly supported that the $J/\psi$ suppression
seen in relativistic heavy ion collision at SPS was a consequence
of QGP(quark gluon plasma) formation\cite{Kharzeev1}.   On the
other hand, other approaches, such as the quark exchange model
\cite{quark-exchange,barnes1}, the meson exchange model
\cite{meson-exchange,Haglin,Lin,Oh1,Ivanov,Sibirtsev,Navarra}, the
QCD sum rule method\cite{Nielsen1,Nielsen2,Nielsen3}, and other
non perturbative methods\cite{Dosch99}, predicted a much larger
cross section of few mbarn. The discrepancy is particulary large
near the threshold region \cite{Lee}.   Although the model
calculations themselves have large uncertainties and model
dependencies,  it is generally believed that such discrepancy
exists because the QCD LO calculation, especially near threshold,
is valid only for a very large quark mass, larger than that of the
bottom quark.  However, no systematic analysis has been worked out
in this context, as the formalism based on the OPE is quite
complicated even in the LO.

A few years ago, Oh, Kim, and Lee \cite{Oh} used Bethe-Salpeter
amplitude and factorization formula to reproduce Peskin's result
on the dissociation cross section.  Because this method is
relatively simple, it opened the possibility to calculate the
higher order correction, which will be carried out in the present
work.

There are two types of initial and final states in the NLO
calculation.  One is $\Phi + q \rightarrow Q + \bar{Q}+ q$, and
the other is $\Phi + g \rightarrow Q + \bar{Q}+ g$, where $\Phi$
is a quarkonium.  In the course of the calculations, collinear
divergence, infrared divergence, and soft-collinear divergence
appear.  Infrared divergence disappear after adding the one loop
diagram with $Q + \bar{Q}$ final state, while the
collinear divergence is eliminated by mass factorization.
Dimensional regularization in the $\overline{MS}$ scheme is used
throughout this work, including the parton distribution function.
The counting scheme, introduced by Peskin, to systematically study
heavy bound states is used and applied to NLO.
In addition, the large $N_c$ limit are taken throughout this work.

In Section 2, the Bethe-salpeter amplitude and the LO calculation
are reviewed. In Section 3, the elementary cross section of $\Phi
+ q \rightarrow Q + \bar{Q}+ q$ is calculated and mass
factorization is introduced. In Section 4, the calculation for the
$\Phi + g \rightarrow Q + \bar{Q}+ g$ process is presented. The
effective four point vertex is introduced in section 5.  It is
then shown how the infrared divergences disappear when the
relevant crossed terms of Born and one loop correction are
included.  Mass factorizations in the process $\Phi + g
\rightarrow Q + \bar{Q}+ g$ are presented separately when the
emitted gluons are hard and soft in section 4 and 5 respectively.
In Section 6, the result is applied to the upsilon dissociation
cross section. Limitations when applied to the charmonium case is
also discussed briefly. Appendix A summarizes the 2 and 3-body
phase space. Appendix B contains the derivation and the list of
angular integration used in the present work.  Appendix C gives
detailed calculations of $\Phi + g \rightarrow Q + \bar{Q}+ g$
diagrams. Appendix D gives some comments about order counting.

\section{LO ($\Phi + g \rightarrow Q + \bar{Q}$) calculation}

Here, the LO result, re-derived by OKL\cite{Oh} using the
Bethe-Salpeter amplitude, is presented again for completeness.

\begin{figure}
\centerline{\includegraphics[width=8cm]{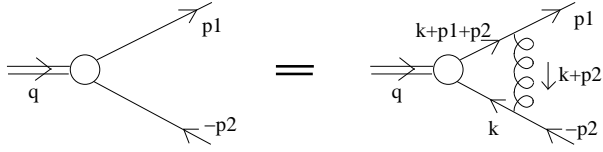}}\caption{The
Bethe-Salpeter equation for quarkonium}\label{fig:BS}
\end{figure}

The Bethe-Salpeter equation, represented diagrammatically in
Fig.~(\ref{fig:BS}), is written as

\begin{eqnarray}
\Gamma_\mu (p_1,-p_2)&=&-ig^2 C_F \int \frac{d^4 K}{(2 \pi)^4}
\gamma^\alpha i\Delta(K+p_1+p_2) \Gamma_\mu
(K+p_1+p_2,K)\nonumber\\
&&\times i\Delta(K) \gamma_\alpha V(K+p_2),
\end{eqnarray}
where $C_F=(N_c^2-1)/N_c$, $i\Delta(K)$ is the quark propagator,
and $i V(K)$ the gluon propagator.

In the heavy quark limit, the $K_0$ contour integration is
dominated by the residue at $k_0=-m-\vec{K}^2/2m+i\varepsilon$
over that at $k_0=-q_0-m-\vec{K}^2/2m+i\varepsilon$. In this
limit, $V(K+p_2)$ becomes three dimensional,
\begin{eqnarray}
V(K+p_2)|_{K_0=-m-\vec{K}^2/2m} \simeq
\frac{1}{|\vec{K}+\vec{p_2}|^2}.
\end{eqnarray}

If further, $\bar{\phi}_\mu(q,\vec{p})$ is defined as,
\begin{eqnarray}
&&\int \frac{p_0}{2\pi}i\Delta(p_1)\Gamma_\mu (p_1,-p_2)
i\Delta(-p_2)\equiv \bar{\phi}_\mu (q,\vec{p}),
\end{eqnarray}
where $q=p_1+p_2$, $p\equiv(p_1-p_2)/2$, the Bethe-Salpeter
equation becomes,
\begin{eqnarray}
\bar{\phi}_\mu(q,\vec{p})&=& i g^2 C_F \int
\frac{dp_0}{2\pi}\Delta(p_1)\bigg[ \int \frac{d^3
K}{(2\pi)^3}\gamma^\alpha \bar{\phi}_\mu (q,\vec{K}+\vec{q}/2)
\gamma_\alpha \frac{1}{|\vec{K}-\vec{p}|^2}\bigg]
\Delta(-p_2)\nonumber\\
&\simeq& -\frac{g^2 C_F}{\epsilon_o+\vec{p}^2/m} \int \frac{d^3
K}{(2\pi)^3}\frac{1+\gamma^0}{2}\gamma^\alpha \bar{\phi}_\mu
(q,\vec{K}) \gamma_\alpha \frac{1-\gamma^0}{2}
\frac{1}{|\vec{K}-\vec{p}|^2},
\end{eqnarray}
in the heavy quark limit and in the $q$ rest frame.
$\epsilon_o=2m-m_\Phi$ is the binding energy of the quarkonium. If
we assume $\bar{\phi}_\mu(q,\vec{p})$ to have a structure of $
\frac{1+\gamma^0}{2}\gamma_i g^i_\mu
\frac{1-\gamma^0}{2}\psi(|\vec{p}|)$, the Bethe-Salpeter equation
reduces to the nonrelativistic Schr$\ddot{o}$dinger equation for
the Coulombic bound state,

\begin{eqnarray}
(\epsilon_o+\vec{p}^2/m)\psi(|\vec{p}|)=  g^2 C_F \int \frac{d^3
K}{(2\pi)^3}\frac{1}{|\vec{K}|^2}\psi(|\vec{K}+\vec{p}|)\label{hardg},
\end{eqnarray}
whose spatial form is,
\begin{eqnarray}
\bigg[ \frac{1}{2\mu} \nabla^2 -\frac{g^2
C_F}{4\pi}\frac{1}{r}\bigg] \psi(r)=\epsilon \psi(r)\label{BS},
\end{eqnarray}
where $\mu=m/2$ is the reduced mass.

At the same time, the Bethe-Salpeter vertex reduces to
\begin{eqnarray}
\Gamma_\mu\bigg(\frac{q}{2}+p,-\frac{q}{2}+p\bigg) &=&
\bigg(\epsilon_o+\frac{\vec{p}^2}{m}\bigg)\psi(|\vec{p}|)\sqrt{\frac{m_\Phi}{N_c}}
\frac{1+\gamma^0}{2}\gamma_i g^i_\mu
\frac{1-\gamma^0}{2}.\label{bsv}
\end{eqnarray}

\begin{figure}
\centerline{\includegraphics[width=12cm]{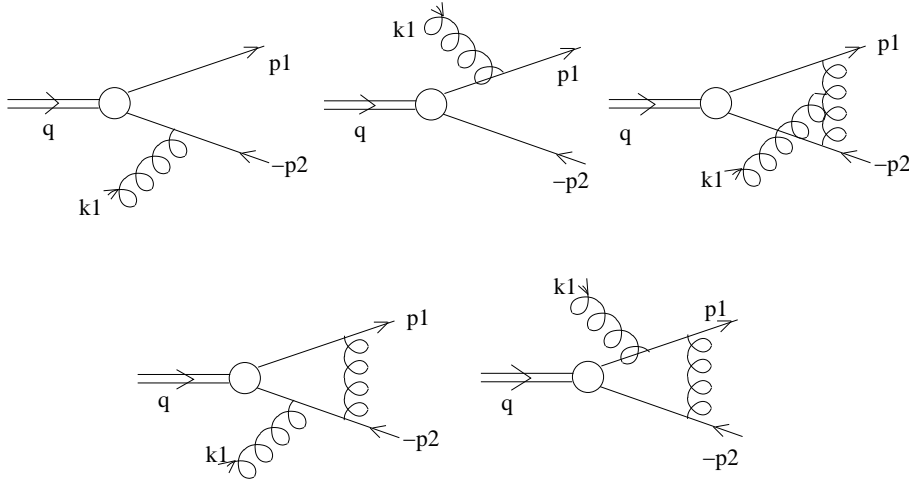}}\caption{Leading
order diagrams for $\Phi+g \rightarrow Q+\bar{Q}$. The two lower
diagrams are suppressed in large $N_c$ limit.}\label{fig:lo}
\end{figure}

Fig.~(\ref{fig:lo}) is the leading order diagrams for quarkonium
dissocation $\Phi + g \rightarrow Q +\bar{Q}$. Among them, the
first three diagrams are of the same order in the large $N_c$
limit, while the last two diagrams are suppressed by $1/N_c$. Such
$N_c$ countings are easily obtained using the double quark line
representation for the gluon lines, since one notes that compared
to the first two diagrams, the third diagram has additionally a
quark loop and two factors of $g$.  This altogether gives a factor
of $N_c g^2$, which scales as $O(1)$ in the large $N_c$ limit, as
the coupling $g $ scales as $1/(N_c)^{\frac{1}{2}}$.  In contrast,
the last two diagrams do not have an additional quark loop while
they have the two additional factors of $g$, and hence are
suppressed by $1/N_c$ compared to the third diagram. Technically,
this can be rephrased in terms of color matrices, as the last two
diagrams carry the color factor $T^b T^a T^b = -T^a/(2N_c)$ and
hence is suppressed by $1/N_c$ compared to third diagram\cite{Oh}.

The counting scheme in scales starts from noting that the binding
energy $\epsilon_o=m(N_c g^2/16\pi)^2 \sim O(mg^4)$.  This
suggests from Eq.~(\ref{hardg}) that the three momenta of the
heavy quarks are of $O(mg^2)$.   The next important step is to
take the external gluon momentum $|\vec{k}|$ and its energy $k^0$
to be of $O(mg^4)$, which are smaller than the typical heavy quark
momentum of $O(mg^2)$ inside the bound state. This is the essence
of the factorization in the present approach; namely, the
separation scale is taken to be of $O(mg^4)$ so that the bound
state property of $O(mg^2)$ can be taken into account as Wilson
coefficients.
Then, from the energy conservation $m_\Phi+k_0=
2m+|\vec{p_1}|^2/2m +|\vec{p_2}|^2/2m$,  one has \cite{peskin},
\begin{eqnarray}
|\vec{p_1}|\sim |\vec{p_2}|\sim O(mg^2), \ \ \ \ \
k^0=|\vec{k}|\sim O(mg^4). \label{counting-scheme}
\end{eqnarray}

Counting in the NLO process $\Phi + q(g)\rightarrow
Q+\bar{Q}+q(g)$ are obtained similarly by assuming that the
incoming and outgoing parton momentum are of $O(mg^4)$.  That is,

\begin{eqnarray}
|\vec{p_1}|\sim |\vec{p_2}|\sim O(mg^2), \ \ \ \ \
k_1^0=|\vec{k_1}|\sim k_2^0=|\vec{k_2}|\sim O(mg^4),\label{gorder}
\end{eqnarray}
where $k_1$ is the incoming quark (gluon) momentum, and $k_2$ is
the outgoing quark (gluon) momentum.  Under this order counting
scheme, quark propagators are expanded as below.

\begin{eqnarray}
\Delta(p+k)&=&\frac{1+\gamma^0}{2}\frac{1}{k_0+i\varepsilon}
+\frac{1+\gamma^0}{2}\frac{\vec{p}\cdot\vec{k}}{m
(k_0+i\varepsilon)^2}+\frac{\not{\vec{p}}}{2mk_0+i\varepsilon}\nonumber\\
\Delta(p-k)&=&\frac{1+\gamma^0}{2}\frac{-1}{k_0-i\varepsilon}
+\frac{1+\gamma^0}{2}\frac{-\vec{p}\cdot\vec{k}}{m
(k_0-i\varepsilon)^2}+\frac{\not{\vec{p}}}{-2mk_0+i\varepsilon}\nonumber\\
\Delta(-p+k)&=&\frac{1-\gamma^0}{2}\frac{-1}{k_0-i\varepsilon}
+\frac{1-\gamma^0}{2}\frac{-\vec{p}\cdot\vec{k}}{m
(k_0-i\varepsilon)^2}+\frac{\not{\vec{p}}}{2mk_0-i\varepsilon}\nonumber\\
\Delta(-p-k)&=&\frac{1-\gamma^0}{2}\frac{1}{k_0+i\varepsilon}
+\frac{1-\gamma^0}{2}\frac{\vec{p}\cdot\vec{k}}{m
(k_0+i\varepsilon)^2}+\frac{-\not{\vec{p}}}{2mk_0+i\varepsilon}\label{qpropa}
\end{eqnarray}
Here, $p$ is the on-shell momentum of heavy quark.  The first term
is of $1/mg^4$ order, and the next two terms are $1/mg^2$ order.
The third diagram of Fig.~(\ref{fig:lo}) seems to be higher order
than previous two diagrams with respect to the coupling $g$. But
they are of same order under above counting scheme. Detail is
given in Appendix D.

Using the Bethe-Salpeter amplitude Eq.~(\ref{bsv}) and the heavy
quark propagators Eq.~(\ref{qpropa}), the leading order amplitude
may be derived as,

\begin{eqnarray}
M^{\mu \nu}_{LO}=-g \sqrt{\frac{m_\Phi}{N_c}} \bigg(
\vec{k_1}\cdot \frac{
\partial \psi ( \vec{p})} {\partial \vec{p}}
g^{\nu 0} + k_{10} \frac{\partial \psi(\vec{p})}{\partial p_j }
g^{\nu j} \bigg) g^{\mu i} \nonumber\\ \times \overline{u}(p_1
)\frac{1+\gamma_0 }{2} \gamma_i \frac{1-\gamma_0}{2} T^a v(p_2 ).
\label{lo}
\end{eqnarray}

The total cross section then becomes
\begin{eqnarray}
\sigma_{\Phi - g}(\lambda)=\frac{2^7
g^2}{3N_c}a_o^2\frac{\bigg(\frac{\lambda}{\epsilon_o}-1\bigg)^{\frac{3}{2}}}
{\bigg(\frac{\lambda}{\epsilon_o}\bigg)^5},\label{locs}
\end{eqnarray}
where $\lambda=q \cdot k_1/m_\Phi$, and the quarkonium is assumed
to be in the coulombic $1S$ state,
\begin{eqnarray}
\nabla \psi_{1s}(\vec{p})=ia_o^{\frac{5}{2}} 32
\sqrt{\pi}\frac{a_o\vec{p}}{(|a_o\vec{p}|^2+1)^3},\label{wavef}
\end{eqnarray}
and $a_o = 16 \pi / (g^2 N_c m)$ is the Bohr radius. The coupling
$g$ and the heavy quark mass $m$ are determined by fitting the
measured quarkonium spectrum, such as $\psi$ and $\psi'$ for the
charmonium states, with those of the coulomb bound states.
Details of the derivations are given in ref. \cite{Oh}.

The hadronic cross section is obtained by folding the partonic
cross section with the parton distribution function, using the
factorization formula,
\begin{eqnarray}
\sigma_{\Phi\ h}(\lambda) = \int^1_0 dx \sigma_{\Phi - parton} (x
\lambda,Q) D_{parton} (x,Q), \label{fac-theorem}
\end{eqnarray}
where $x$ is the momentum fraction of a parton, and $Q$ is the
separation scale.   As has been stated earlier, the scale $Q$ is
set to be the binding energy, which is of $O(mg^4)$, then it is
natural to include the `bound state' properties obtained from
momentum scale of $O(mg^2)$ in the Wilson coefficient.

At this point, it should be noted that the factorization theorem
in Eq.(\ref{fac-theorem}) is valid only when the mass of the quark
is very large, so that the binding energy of  $O(mg^4)$ becomes
larger than the typical hard scale $> 1$ GeV.  Otherwise, the two
standard sets of corrections will not be small.  Namely, the
higher order correction to the perturbative cross section will not
converge and the higher twist effects will not be negligible.

If the heavy quark is not sufficiently heavy, the time scale
involved in forming the bound state will not be short enough
compared to the typical time scale involved for the bound state to
interact with the external partons, and hence, the contributions
from multiple gluonic interactions will not be negligible. Such
multiple gluonic effects correspond to the higher twist effects.
Moreover, even if one calculates the Wilson coefficients for such
higher twist effects, nothing much is known about the higher twist
distribution functions inside the hadrons, and the corresponding
contribution to the hadronic cross sections can not be calculated.

The perturbative cross section of the leading twist also has its
own problems when the quark mass is not heavy enough.  To begin
with, to implement Eq.(\ref{fac-theorem}) at the separation scale
$Q=\epsilon_o$, one needs the parton distribution function
$D_{parton}(x,\epsilon)$ defined at that scale.  Moreover, the
perturbative calculation for the leading order cross section
$\sigma_{\Phi-parton}(x\lambda,\epsilon)$ may not be convergent.
Nonetheless, all such questions can be answered by the explicit
NLO calculation, which will provide an quantitative estimate of
the correction to the LO cross section, and also determine the
valid energy range of the LO cross section.

The binding energies for both the charmonium and the Upsilon
systems are around 0.75 GeV, and the aforementioned corrections
are potentially not small.  However, by applying our formal NLO
calculation to the Upsilon system, we will investigate the
convergence and the valid energy range  of the leading twist cross
section.   As for the parton distribution function, we will use
the MRST parton distribution function at its minimum scale
$Q^2=1.25$ GeV$^2$, and investigate its uncertainty due to the
variation in the scale to larger values.

\section{NLO process $\Phi + q \rightarrow Q + \bar{Q}+ q$}

\subsection{Collinear divergent elementary cross section}

\begin{figure}
\centerline{\includegraphics[width=12cm]{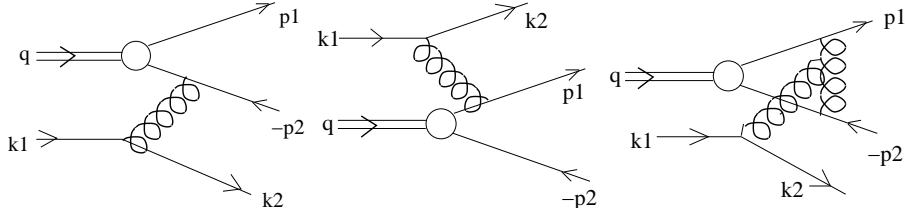}}\caption{Next
to leading order diagram with initial quark $\Phi+q \rightarrow
Q+\bar{Q}+q$}. \label{fig:nloq}
\end{figure}

Fig~(\ref{fig:nloq}) represents the lowest order diagrams
involving the quarks,  $\Phi + q \rightarrow Q +\bar{Q}+q$.  The
invariant matrix element for this process is given by
\begin{eqnarray}
M^\mu_{NLO-q} &=& g \overline{u}(k_2 ) \gamma_\nu T^a u(k_1 )
\frac{1}{(k_1- k_2 )^2 } \nonumber\\
&&\times -g \sqrt{\frac{m_\Phi}{N_c}} \bigg[
(\vec{k_1}-\vec{k_2})\cdot \frac{
\partial \psi ( \vec{p})} {\partial \vec{k}}
g^{\nu 0} + (k_{10} -k_{20} ) \frac{\psi(\vec{p})}{\partial p_j }
g_j^\nu \bigg]  \nonumber\\
&&\times \overline{u}(p_1 )\frac{1+\gamma_0 }{2} \gamma_i g^{\mu
i} \frac{1-\gamma_0}{2} T^a v(p_2 ).
\end{eqnarray}

The averaged square in D-dimension is
\begin{eqnarray}
\overline{|M|}^2_{NLO-q} &=& \frac{2^{12}}{3}\pi g^4 m^2 m_\Phi
\bigg( \frac{ \epsilon_o }{m} \bigg)^{\frac{5}{2}}
\frac{k_{10}-k_{20}-\epsilon_o}{(k_{10}-k_{20})^6}\nonumber\\
&&\times \bigg[ -\frac{1}{2} +\frac{k_{10}^2 +k_{20}^2 }{2k_1
\cdot k_2} +(D-4)\frac{(k_{10}-k_{20})^2}{4 k_1 \cdot k_2}\bigg].
\end{eqnarray}

It may be modified to the following covariant form,
\begin{eqnarray}
\overline{|M|}_{NLO-q}^2  &=& \frac{2^{11}}{3}\pi g^4 m^2
(2m_\Phi)^6 \bigg( \frac{ \epsilon_o }{m} \bigg)^{\frac{5}{2}}
\frac{ \acute{s}+\acute{u}- 2 m_\Phi
\epsilon_o}{(\acute{s}+\acute{u})^6}
\nonumber\\
&& \times \bigg[-\frac{1}{2} - \frac{\acute{s}^2 + \acute{u}^2
}{(2 m_\Phi )^2
\acute{t}}-(D-4)\frac{(\acute{s}+\acute{u})^2}{2\acute{t}} \bigg],
\label{inv}
\end{eqnarray}
where $\acute{s}\equiv 2q \cdot k_1$, $ \acute{u}\equiv -2q \cdot
k_2$, and $\acute{t}\equiv -2k_1 \cdot k_2$.

The parameterization of the three body phase space follows
ref.\cite{Beenakker}.   The initial and final momenta are set to
the following,

\begin{eqnarray}
q&=&(E_q, 0, ... , 0, 0,|\vec{p}| \sin\varphi,|\vec{p}|\cos\varphi-k_{10})\nonumber\\
k_1&=&(k_{10}, 0, ... , 0, 0, 0,k_{10}) \nonumber\\
p_1&=&(E_1, 0, ... , 0,
-k_{20}\sin\theta_1\sin\theta_2,-k_{20}\sin\theta_1\cos\theta_2,-k_{20}\cos\theta_1)\nonumber\\
p_2&=&(E_2, 0, ... , 0, 0, |\vec{p}|\sin\varphi,
|\vec{p}|\cos\varphi)\nonumber\\
k_2&=&(k_{20}, 0, ... ,
0,k_{20}\sin\theta_1\sin\theta_2,k_{20}\sin\theta_1\cos\theta_2,k_{20}\cos\theta_1).
\end{eqnarray}

The following new invariant variables are introduced,

\begin{eqnarray}
t_1 &\equiv& (k_1-p_2)^2-m^2=-2k_1 \cdot p_2 \nonumber\\
u_1 &\equiv& (q-p_2)^2-m_\Phi^2-m^2=-2q \cdot p_2 \nonumber\\
s_4 &\equiv& (k_2+p_1)^2-m^2 =2k_2 \cdot p_1=s+t_1+u_1,
\end{eqnarray}
where $s=(q+k_1)^2$.

$E_q, k_{10}, E_1 ... $ may be expressed in terms of the invariant
variables $s(\acute{s}),t_1, u_1, s_4$.

\begin{eqnarray}
E_q&=&\frac{s+u_1+m_\Phi^2}{2\sqrt{s_4+m^2}} \nonumber\\
k_{10}&=&\frac{s+t_1-m_\Phi^2}{2\sqrt{s_4+m^2}} \nonumber\\
E_1&=&\frac{s_4+2m^2}{2\sqrt{s_4+m^2}}\nonumber\\
E_2&=&-\frac{t_1+u_1+2m^2}{2\sqrt{s_4+m^2}}\nonumber\\
k_{20}&=&\frac{s_4}{2\sqrt{s_4+m^2}}\ \nonumber\\
|\vec{p}|&=&\frac{\sqrt{(u_1+t_1)^2-4m^2
s}}{2\sqrt{s_4+m^2}}\nonumber\\
\cos \varphi&=&\frac{t_1 s_4
-s(u_1+2m^2)+(s_4-s+2m^2)m_\Phi^2}{(s+t_1-m_\Phi^2)\sqrt{(t_1+u_1)^2-4m^2
s }}. \label{kineticvariable}
\end{eqnarray}

Using these relations, Eq.~(\ref{inv}) are expressed by five
variables $s(\acute{s}),t_1, u_1, \theta_1, \theta_2$, because

\begin{eqnarray}
\acute{u}&=&(q-k_2)^2-m_\Phi^2=2k_{20}(-E_q+(|\vec{p}|\cos\varphi-k_{10})\cos\theta_1
+|\vec{p}|\sin\varphi\sin\theta_1
\cos\theta_2)\nonumber\\
\acute{t}&=&(k_1-k_2)^2=-2k_{10}k_{20}(1-\cos\theta_1).
\end{eqnarray}

The differential cross section for the three body decay is

\begin{eqnarray}
\acute{s}^2 \frac{d^2 \sigma }{dt_1 du_1} &=&
\frac{1}{2}\frac{1}{(4\pi)^D}\frac{\mu^{-D+4}}{\Gamma(D-3)}\bigg(\frac{\acute{s}u_1
t_1-m_\Phi^2 t_1^2-m^2\acute{s}^2}{\acute{s}^2 \mu^2 }
\bigg)^{\frac{D-4}{2}}
\frac{s_4^{D-3}}{(s_4+m^2)^{\frac{D}{2}-1}}\nonumber\\
&& \times \int^\pi_0 d\theta_1 \sin^{D-3}\theta_1 \int^\pi_0
d\theta_2 \sin^{D-4}\theta_2 \overline{|M|}^2.
\end{eqnarray}

The derivation is given in Appendix A. When $\theta_1=0$,
${\acute{t}}=0$ and the term $1/{\acute{t}}$ in Eq.~(\ref{inv})
gives collinear divergence. Defining $I^{(i,j)}$ as below,

\begin{eqnarray}
I^{(i,j)} \equiv \int^\pi_0 d\theta_1 \sin^{D-3}\theta_1
\int^\pi_0 d\theta_2 \sin^{D-4}\theta_2
\frac{1}{\acute{t}^i(\acute{u}+\acute{s})^j},
\end{eqnarray}
and expanding it with respect to $D-4$

\begin{eqnarray}
I^{(1,j)}=\frac{1}{D-4}I^{(1,j)}_{-1} +I^{(1,j)}_0 +O(D-4),
\end{eqnarray}
the differential cross section of $\Phi + q \rightarrow Q +
\bar{Q}+ q$ is regularized as below,

\begin{eqnarray}
\acute{s}^2 \frac{d^2 \sigma_{NLO-q}}{dt_1 du_1} && =
\frac{2^8}{3}\frac{1}{(4\pi)^3}g^4 m^2 (2m_\Phi)^4
\bigg(\frac{\epsilon_o}{m}\bigg)^{\frac{5}{2}}\frac{s_4}{s_4+m^2}\bigg[\nonumber\\
&& \frac{1}{D-4}\bigg( -I^{(1,3)}_{-1}+2(\acute{s}+
m_\Phi\epsilon_o)I^{(1,4)}_{-1}-2\acute{s}(\acute{s}+2m_\Phi\epsilon_o)I^{(1,5)}_{-1}
+4\acute{s}^2m_\Phi\epsilon_o
I^{(1,6)}_{-1}\bigg)\nonumber\\
&&-2m_\Phi^2I^{(0,5)}+4m_\Phi^3\epsilon_o
I^{(0,6)}-I^{(1,3)}_{0}+2(\acute{s}+
m_\Phi\epsilon_o)I^{(1,4)}_{0}\nonumber\\
&&-2\acute{s}(\acute{s}+2m_\Phi\epsilon_o)I^{(1,5)}_{0}
+4\acute{s}^2m_\Phi\epsilon_o I^{(1,6)}_{0}\nonumber\\
&&+\bigg(\gamma_E+\ln \frac{s_4 \sqrt{ \acute{s}u_1 t_1-m_\Phi^2
t_1^2-m^2\acute{s}^2}}{4\pi\acute{s} \mu^2 \sqrt{s_4+m^2}}
\bigg)\nonumber\\
&&\ \ \ \ \times\bigg( -I^{(1,3)}_{-1}+2(\acute{s}+
m_\Phi\epsilon_o)I^{(1,4)}_{-1}
-2\acute{s}(\acute{s}+2m_\Phi\epsilon_o)I^{(1,5)}_{-1}+4\acute{s}^2m_\Phi\epsilon_o
I^{(1,6)}_{-1}\bigg) \nonumber\\
&& -\frac{1}{2}I_{-1}^{(1,3)}+m_\Phi \epsilon_o I_{-1}^{(1,4)}\ \
\ \ \bigg].
\end{eqnarray}

From the list of integration in Appendix B,

\begin{eqnarray}
I^{(1,j)}_{-1}=\frac{2\pi}{a} \frac{1}{(A+B)^j}=\frac{2\pi}{a}
\frac{1}{(\acute{s}X)^j},
\end{eqnarray}
where $X = -(u_1+m_\Phi^2)/(\acute{s}+t_1)$, which is
$1-k_{20}/k_{10}$ in the quarkonium rest frame.  The definition of
$a$, $A$, and $B$ are given in Appendix B.  The terms proportional
to $1/(D-4)$ come from collinear divergence, and are eliminated by
mass factorization.

\subsection{Mass factorization}

\begin{figure}
\centerline{\includegraphics[width=8cm]{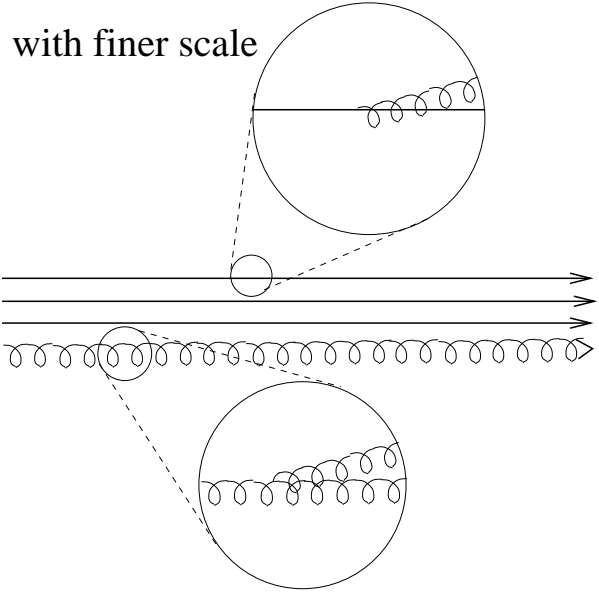}} \caption{Scale
dependence of parton distribution function.} \label{fig:scale}
\end{figure}

Collinear divergence is eliminated by mass factorization, which
moves the divergent contribution to the parton distribution
function. When one parton is seen with finer scale, it is not seen
as one parton, but a sum of several collinear partons as shown in
Fig.~(\ref{fig:scale}). In other words, parton distribution
function has scale dependence, and
the collinear parton with transverse momentum less than the scale
should be included in the parton distribution function.  Only the
parton with larger transverse momentum is included  in the
perturbative calculations. Therefore collinear partons should be
subtracted from the perturbative calculation. In the
$\overline{MS}$ scheme, mass factorization is defined as

\begin{eqnarray}
\acute{s}^2 \frac{d^2 \hat{\sigma}_{NLO-i}}{d t_1
du_1}=\acute{s}^2 \frac{d^2 \sigma_{NLO-i}}{d t_1
du_1}-\frac{\alpha_s}{2 \pi} \int^1_0 \frac{dx}{x} P_{ji}(x)
\bigg( \frac{2}{D-4}+\gamma_E+\ln\frac{Q^2}{4\pi\mu^2} \bigg)
\hat{\acute{s}}^2 \frac{d^2 \hat{\sigma}_{LO-j}}{d\hat{t_1}du_1},
\end{eqnarray}
where $\hat{\acute{s}}=x\acute{s}$, $\hat{t_1}= xt_1$. The
parenthesis means the integration of transverse momentum from zero
to momentum scale $Q$. $\sigma_{NLO-i}$ is the next leading order
cross section of quarkonium and parton $i$. $\hat{\sigma}_{NLO-i}$
is the reduced cross section after mass factorization, which is
finite. $\sigma_{LO-j}$ is the leading order cross section of
quarkonium and parton $j$, and
$\hat{\sigma}_{LO-j}=\sigma_{LO-j}$. $P_{ji}$ is the splitting
function of parton i to parton j. In $\Phi+q \rightarrow
Q+\bar{Q}+q$ process,

\begin{eqnarray}
P_{gq}(x)=\frac{N_c^2-1}{2N_c} \bigg[\frac{1+(1-x)^2}{x}\bigg]
\end{eqnarray}
is needed. The LO differential cross section is

\begin{eqnarray}
\hat{\acute{s}}^2 \frac{d^2
\hat{\sigma}_{LO-j}}{d\hat{t_1}du_1}&=&\frac{2^7}{3 N_c} g^2 m^2
(2m_\Phi)^4\bigg(\frac{\epsilon_o}{m}\bigg)^{\frac{5}{2}}\frac{\acute{s}x-2m_\Phi
\epsilon_o}{(
\acute{s}x)^4}\delta(\acute{s}x+t_1x+u_1+m_\Phi^2)\nonumber\\
&&\times
\frac{D-2}{2}\frac{1}{\Gamma(D/2-1)}\bigg(\frac{\acute{s}u_1
t_1-m_\Phi^2 t_1^2-m^2\acute{s}^2}{4\pi \mu^2
\acute{s}^2}\bigg)^{\frac{D-4}{2}}.
\end{eqnarray}
The LO differential cross section can be obtained from
substituting Eq.~(\ref{lo}) into Eq.~(\ref{2-body}) in Appendix A.

After mass factorization, the reduced differential cross section
is

\begin{eqnarray}
\acute{s}^2 \frac{d^2 \hat{\sigma}_{NLO-q}}{d t_1
du_1}&=&\frac{2^7}{3}\frac{1}{(4\pi)^2}g^4 m^2
(2m_\Phi)^4\bigg(\frac{\epsilon_o}{m}\bigg)^{\frac{5}{2}}\frac{1}{\acute{s}+t_1}\nonumber\\
&\times& \bigg[
\frac{m_\Phi^2(\acute{s}+t_1)}{\pi}\frac{s_4}{s_4+m^2}(-I^{(0,5)}+2m_\Phi\epsilon_o
I^{(0,6)})\nonumber\\
&&+\frac{1}{(\acute{s}X)^3}\bigg(\ln\frac{s_4^2}{Q^2(s_4+m^2)}+I^{(1,3)*}_0
\bigg)\nonumber\\
&&-\frac{2(\acute{s}+m_\Phi \epsilon_o)
}{(\acute{s}X)^4}\bigg(\ln\frac{s_4^2}{Q^2(s_4+m^2)}+I^{(1,4)*}_0
\bigg)+\frac{2\acute{s}}{(\acute{s}X)^4}\nonumber\\
&&+\frac{2\acute{s}(\acute{s}+2m_\Phi\epsilon_o)}{(\acute{s}X)^5}
\bigg(\ln\frac{s_4^2}{Q^2(s_4+m^2)}+I^{(1,5)*}_0-1\bigg)\nonumber\\
&&-\frac{4m_\Phi \epsilon_o \acute{s}^2
}{(\acute{s}X)^6}\bigg(\ln\frac{s_4^2}{Q^2(s_4+m^2)}+I^{(1,6)*}_0-1
\bigg)\ \ \ \ \ \bigg]\label{nlo-q},
\end{eqnarray}
where $I^{(1,j)*}_0 \equiv a(A+B)^j I^{(1,j)}_0/{\pi} =
a(\acute{s}X)^j I^{(1,j)}_0/{\pi}$.

One should also note that the cross sections to NLO in
Eq.~(\ref{nlo-q}) and to LO in Eq.~(\ref{locs}) have the same
large $N_c$ scaling, as the coupling constant $g$ scales as
$1/\sqrt{N_c}$.

The threshold of differential cross section in (\ref{nlo-q}) is
$\acute{s}=2m_\Phi \epsilon_o$. This comes from the quarkonium
wavefunction (\ref{wavef}). However, the physical threshold is
$\acute{s}=2m_\Phi \epsilon_o+\epsilon_o^2$. The term
$\epsilon_o^2$ was ignored because $g$ is of $O(mg^4)$. We
circumvent this problem by substituting
$\epsilon_o+\epsilon_o^2/2m_\Phi$ for $\epsilon_o$ in the
differential cross section.

\subsection{Dalitz plot}

\begin{figure}
\centerline{\includegraphics[width=8cm]{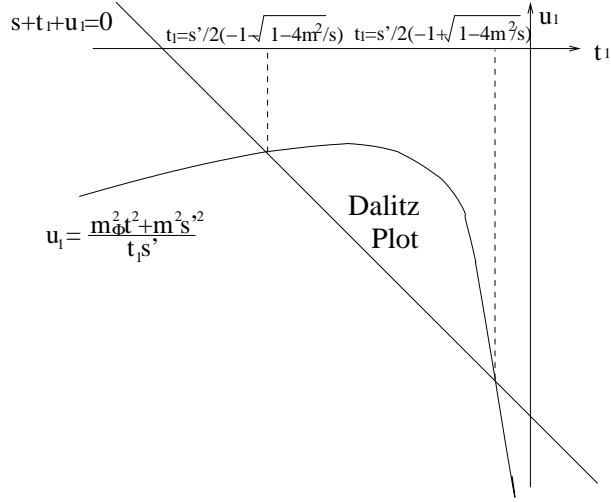}}\caption{Dalitz
plot for $\Phi+q\rightarrow Q+\bar{Q}+q$.} \label{fig:Dalitz1}
\end{figure}

Fig.~(\ref{fig:Dalitz1}) is the Dalitz plot, which is drawn under
the following two conditions.

\begin{eqnarray}
s_4&=&\acute{s}+u_1+t_1+m_\Phi^2\geq 0 \\
|\cos\chi|^2&=&\bigg|\frac{-2u_1
s+(s+m_\Phi^2)(t_1+u_1)}{\acute{s}\sqrt{(t_1+u_1)^2-4sm^2}}\bigg|^2\leq
1,
\end{eqnarray}
here $\chi$ is the angle between $\vec{q}$ and $\vec{p}_2$, or
$\vec{q}$ and $\vec{p}_1+\vec{k}_2$ in CM frame Eq.~(\ref{angle}).
The elementary total cross section is obtained by numerically
integrating Eq.~(\ref{nlo-q}) over the Dalitz plot.  Furthermore,
the hadronic cross section is obtained by folding it with the
quark distribution function.

\section{Hard part of the $\Phi + g \rightarrow Q + \bar{Q}+ g$ process}

\subsection{Soft and/or collinear divergent elementary cross
section}

\begin{figure}
\centerline{
\includegraphics[width=8cm]{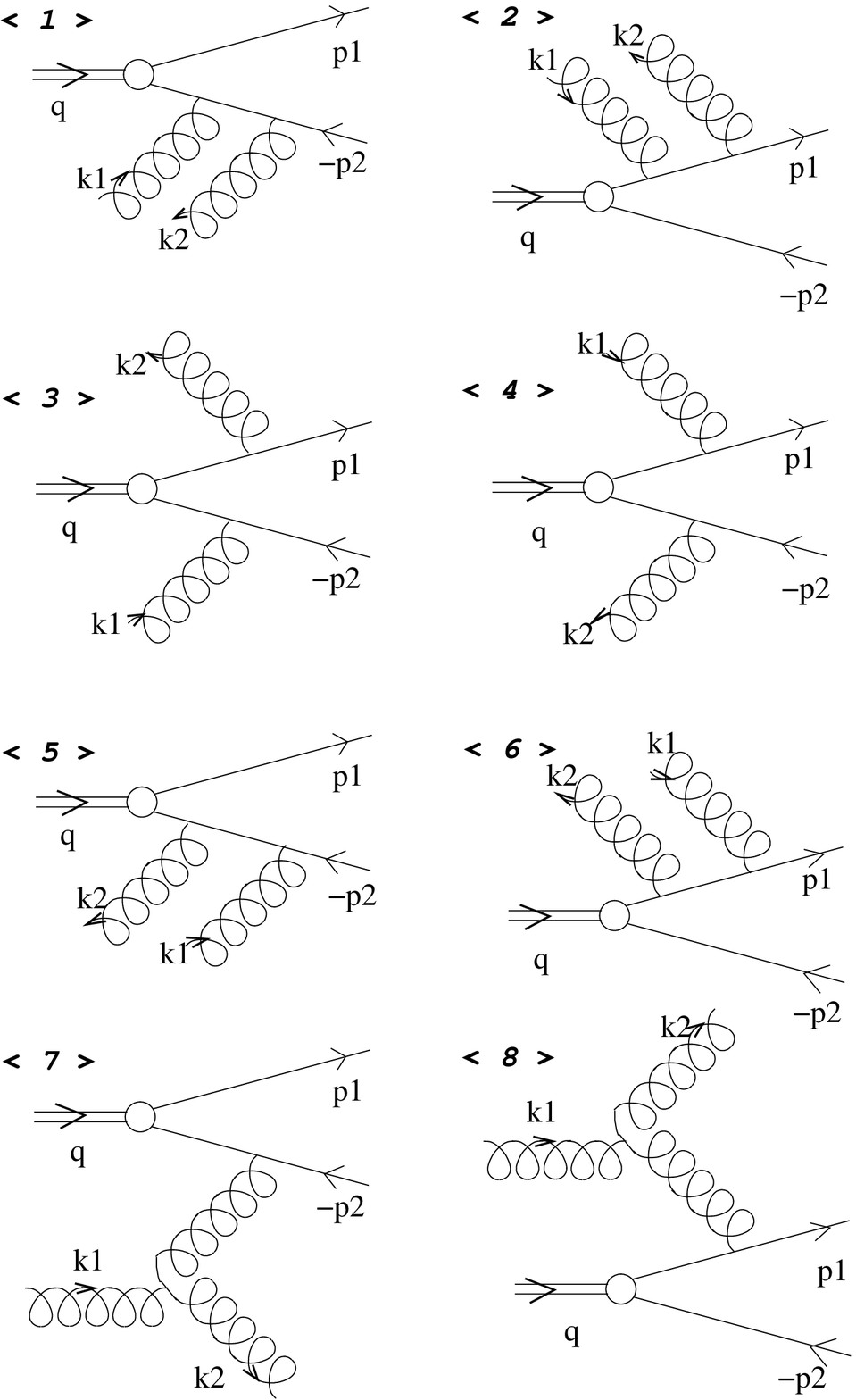}\hfill
\includegraphics[width=8cm]{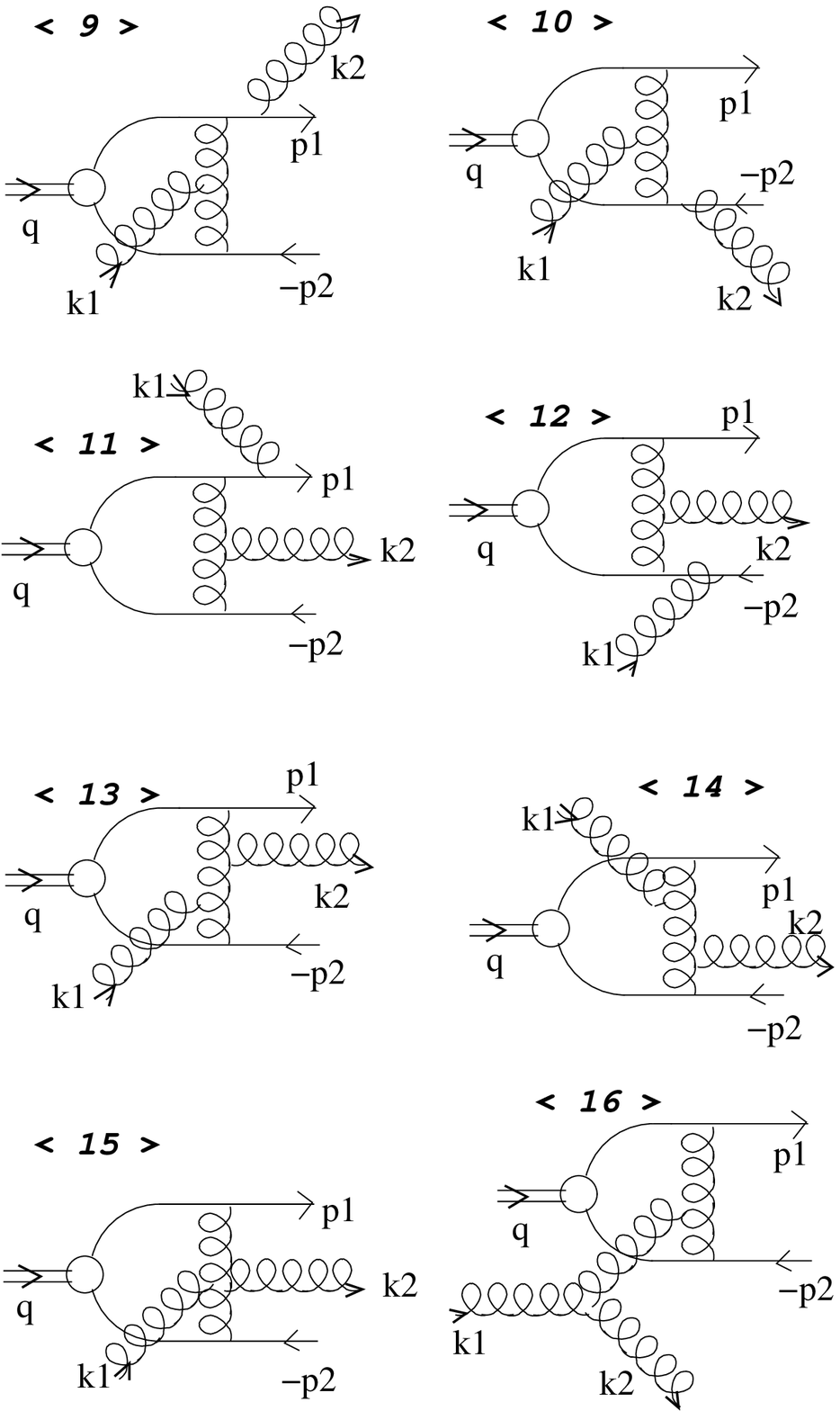}}\caption{Next to leading
order diagram with initial gluon $\Phi+g \rightarrow
Q+\bar{Q}+g$.} \label{fig:nlog}
\end{figure}

Fig.~(\ref{fig:nlog}) represent the diagrams for the process $\Phi
+ g \rightarrow Q + \bar{Q}+ g$.  Among them, diagrams (13), (14),
and (15) are ignored because they are higher order in $g$ compared
to the rest of the diagrams, in the present counting scheme, where
the momentum of the internal gluon, which binds heavy quark and
antiquark, is of $O(mg^2)$ (from Eq.~(\ref{hardg})), and that of
the external gluon is of $O(mg^4)$ (from Eq.~(\ref{gorder})).
Details are given in Appendix D.

The invariant amplitude in quarkonium rest frame is

\begin{eqnarray}
M^{\mu \nu \lambda (a,b) }_{NLO-g} &=& \bigg[ \bigg(\frac{
\partial \psi (p )}{\partial \vec{p}} \cdot \vec{k_1} \bigg)
\bigg(-g^\lambda_0 g^\nu_0 \frac{1}{k_{20}} + \frac{1}{k_1 \cdot
k_2 } (g^\lambda_0 k^\nu_2
+g^\nu_0 k^\lambda_1 - g^{\nu \lambda } k_{20}) \bigg) \nonumber\\
&&+\bigg(\frac{ \partial \psi (p )}{\partial \vec{p}} \cdot
\vec{k_2} \bigg) \bigg(g^\lambda_0 g^\nu_0 \frac{1}{k_{10}} -
\frac{1}{k_1 \cdot k_2 } (g^\lambda_0 k^\nu_2 +g^\nu_0 k^\lambda_1
- g^{\nu \lambda } k_{10}) \bigg)
\nonumber\\
&&+(k_{10 } -k_{20 }) \frac{\psi(p) }{\partial p_{j}}
\bigg(-g^\lambda_j g^\nu_0 \frac{1}{k_{10 }} -g^\lambda_0 g^\nu_j
\frac{1}{k_{20}} +\frac{1}{k_1 \cdot k_2 } (g^\nu_j k^\lambda_1
+g^\lambda_j k^\nu_2 )\bigg) \bigg] \nonumber\\
&&\ \ \ \ \ \ \times g^2 \sqrt{\frac{m_\Phi}{N_c }}
\overline{u}(p_1 )\frac{1+\gamma_0 }{2} \gamma_i g^{\mu
i}\frac{1-\gamma_0}{2}\ [T^a ,T^b ]\ v(p_2 ). \label{amp-g}
\end{eqnarray}

The detailed derivation of this result is given in Appendix C.  It
was checked that the following current conservation conditions are
satisfied.
\begin{eqnarray}
q_\mu M^{\mu \nu \lambda}_{NLO-g} = k_{1 \nu} M^{\mu \nu
\lambda}_{NLO-g} = k_{2 \lambda} M^{\mu \nu
\lambda}_{NLO-g}=0.\nonumber
\end{eqnarray}

The averaged squared amplitude is

\begin{eqnarray}
\overline{|M|}_{NLO-g}^2  =\frac{2^{11}}{3} \pi g^4 m^2 (2 m_\Phi
)^6 \bigg(\frac{\epsilon_o }{m}\bigg)^{\frac{5}{2}} \bigg(
\frac{1}{(\acute{s}+\acute{u})^5 } -\frac{2m_\Phi
\epsilon_o}{(\acute{s}+\acute{u})^6 } \bigg) \bigg[ \frac{(2m_\Phi
)^2 }{2}
\frac{\acute{t}}{\acute{s} \acute{u}}\nonumber\\
-\frac{D-2}{2}\bigg( \frac{2 \acute{s}}{\acute{u}}
+\frac{2\acute{u}}{\acute{s}}+\frac{\acute{u}^2}{\acute{s}^2}
+\frac{\acute{s}^2}{\acute{u}^2}\bigg)-D\nonumber\\
+\frac{D-2}{2m^2_\Phi \acute{t}} \bigg( \frac{(\acute{s}^2
+\acute{u}^2 )^2}{\acute{s} \acute{u}} +2\acute{s}^2 +2\acute{u}^2
+\acute{s}\acute{u} \bigg) \bigg]. \label{ave}
\end{eqnarray}

The first line is of order $\acute{t}$, the second line  of order
$\acute{t}^0$, and the third line of order $\acute{t}^{-1}$.

Differential cross section from the first line is

\begin{eqnarray}
\frac{2^{7}}{3}\frac{1}{(4 \pi)^3} g^4 m^2 (2 m_\Phi )^8
(\frac{\epsilon_o
}{m})^{\frac{5}{2}}\frac{s_4}{s_4+m^2}\bigg[\bigg(1-\frac{2m_\Phi
\epsilon_o
}{\acute{s}}\bigg)\bigg(\frac{J^{(-1,1)}}{\acute{s}^6}-\frac{I^{(-1,1)}}{\acute{s}^6}
\nonumber\\-\frac{I^{(-1,2)}}{\acute{s}^5}-\frac{I^{(-1,3)}}{\acute{s}^4}
-\frac{I^{(-1,4)}}{\acute{s}^3}
-\frac{I^{(-1,5)}}{\acute{s}^2}\bigg)+\frac{2m_\Phi\epsilon_o
}{\acute{s}^2}I^{(-1,6)}  \bigg],\label{hard1}
\end{eqnarray}
where $J^{(i,j)}$ is defined as follows,

\begin{eqnarray}
J^{(i,j)}&\equiv& \int^\pi_0 d\theta_1 \sin^{D-3}\theta_1
\int^\pi_0 d\theta_2
\sin^{D-4}\theta_2\frac{1}{(\acute{t})^i(\acute{u})^j}\nonumber\\
&=&\int^\pi_0 d\theta_1  \int^\pi_0 d\theta_2
\frac{\sin^{D-3}\theta_1 \sin^{D-4}\theta_2
}{(a-a\cos\theta_1)^i(\acute{A}+\acute{B}\cos\theta_1+\acute{C}\sin\theta_1
\cos\theta_2)^j},\nonumber
\end{eqnarray}
and, $\acute{A}=-2E_q k_{20}$, $\acute{B}=B$, and $\acute{C}=C$.
The products of invariant variables are decomposed as below.

\begin{eqnarray}
\frac{1}{(\acute{s}+\acute{u})^5
\acute{u}}&=&\frac{1}{\acute{u}\acute{s}^5
}-\frac{1}{(\acute{s}+\acute{u})
\acute{s}^5}-\frac{1}{(\acute{s}+\acute{u})^2
\acute{s}^4}-\frac{1}{(\acute{s}+\acute{u})^3
\acute{s}^3}-\frac{1}{(\acute{s}+\acute{u})^4
\acute{s}^2}\nonumber\\
&&-\frac{1}{(\acute{s}+\acute{u})^5 \acute{s}}
\end{eqnarray}

The first line has no divergent term.

Differential cross section from the second line is

\begin{eqnarray}
&&\frac{2^{8}}{3}\frac{1}{(4\pi)^3} g^4 m^2 (2 m_\Phi )^6
\bigg(\frac{\epsilon_o }{m}\bigg)^{\frac{5}{2}}\frac{s_4}{s_4+m^2}
\bigg[J^{(0,1)}\bigg(\frac{3\acute{s}-8m_\Phi\epsilon_o}{\acute{s}^5}\bigg)\nonumber\\
&&+ \frac{D-2}{2\Gamma(D-3)}\bigg(\frac{s_4 \sqrt{\acute{s}u_1
t_1-m_\Phi^2 t_1^2-m^2\acute{s}^2}}{4\pi \acute{s}\mu^2
\sqrt{s_4+m^2}}
\bigg)^{D-4}J^{(0,2)}\bigg(\frac{-\acute{s}+2m_\Phi\epsilon_o}{\acute{s}^4}\bigg)\nonumber\\
&&+
I^{(0,1)}\bigg(\frac{-3\acute{s}+8m_\Phi\epsilon_o}{\acute{s}^5}\bigg)+
I^{(0,2)}\bigg(\frac{-2\acute{s}+6m_\Phi\epsilon_o}{\acute{s}^4}\bigg)\nonumber\\&&+
I^{(0,3)}\bigg(\frac{-2\acute{s}+4m_\Phi\epsilon_o}{\acute{s}^3}\bigg)+
\frac{4m_\Phi\epsilon_o}{\acute{s}^2}I^{(0,4)}-2I^{(0,5)} +4m_\Phi
\epsilon_o I^{(0,6)} \bigg].\label{hard2}
\end{eqnarray}
$J^{(0,2)}$ is the soft divergent term, because it is proportional
to $1/\acute{u}^2$, which is $1/k_{20}^2$ in quarkonium rest
frame.

Differential cross section from the third line is

\begin{eqnarray}
&&\frac{2^{10}}{3}\frac{1}{(4\pi)^3} g^4 m^2 (2 m_\Phi )^4
\bigg(\frac{\epsilon_o }{m}\bigg)^{\frac{5}{2}}
\frac{s_4}{s_4+m^2}\nonumber\\
&&\times \bigg[
\bigg(\frac{1}{D-4}+\gamma_E+\ln\frac{s_4\sqrt{\acute{s}u_1 t_1
-m_\Phi^2 t_1^2
-m^2\acute{s}^2}}{4\pi\mu^2\acute{s}\sqrt{s_4+m^2}}+\frac{1}{2}\ \bigg)\nonumber\\
&&\ \ \ \ \ \ \times\frac{2 \pi}{a} \bigg(
\frac{\acute{s}-2m_\Phi\epsilon_o}{\acute{s}^3}\frac{1}{\acute{s}(x-1)}
-\frac{\acute{s}-2m_\Phi\epsilon_o}{\acute{s}^3}\frac{1}{\acute{s}x}
+\frac{2m_\Phi \epsilon_o}{\acute{s}^2} \frac{1}{(\acute{s}x)^2}\nonumber\\
&&-2\frac{1}{(\acute{s}x)^3} +(\acute{s}+4m_\Phi \epsilon_o
)\frac{1}{(\acute{s}x)^4}
-\acute{s}(\acute{s}+2m_\Phi\epsilon_o)\frac{1}{(\acute{s}x)^5}+2m_\Phi\epsilon_o
\acute{s}^2 \frac{1}{(\acute{s}x)^6} \bigg) \nonumber\\
&&\ \ \ \ \ \
+\frac{\acute{s}-2m_\Phi\epsilon_o}{\acute{s}^3}J^{(1,1)}_0
-\frac{\acute{s}-2m_\Phi\epsilon_o}{\acute{s}^3}I^{(1,1)}_0
+\frac{2m_\Phi \epsilon_o}{\acute{s}^2} I^{(1,2)}_0 -2I^{(1,3)}_0\nonumber\\
&&\ \ \ \ \ \ +(\acute{s}+4m_\Phi \epsilon_o )I^{(1,4)}_0
-\acute{s}(\acute{s}+2m_\Phi\epsilon_o)I^{(1,5)}_0
+2m_\Phi\epsilon_o\acute{s}^2 I^{(1,6)}_0 \bigg],\label{hard3}
\end{eqnarray}
where $J^{(1,j)}$ is expanded with respect to $D-4$

\begin{eqnarray}
J^{(1,j)} = \frac{1}{D-4}J^{(1,j)}_{-1} +J^{(1,j)}_0 +O(D-4),
\end{eqnarray}
and,

\begin{eqnarray}
J^{(1,j)}_{-1}&=&\frac{2\pi}{a}
\frac{1}{(\acute{A}+\acute{B})^j}=\frac{2\pi}{a}
\frac{1}{(\acute{s}(X-1))^j}.
\end{eqnarray}

$1/(D-4)$ term of Eq.~(\ref{hard3}) is collinear divergent.
Additionally $1/{a(1-X)}$ term gives soft divergence, because both
$a$ and $1-X$ are proportional to $k_{20}$. Thus this term gives
soft-collinear divergence.

\subsection{Mass factorization in hard gluon emitted region}

\begin{figure}
\centerline{\includegraphics[width=8cm]{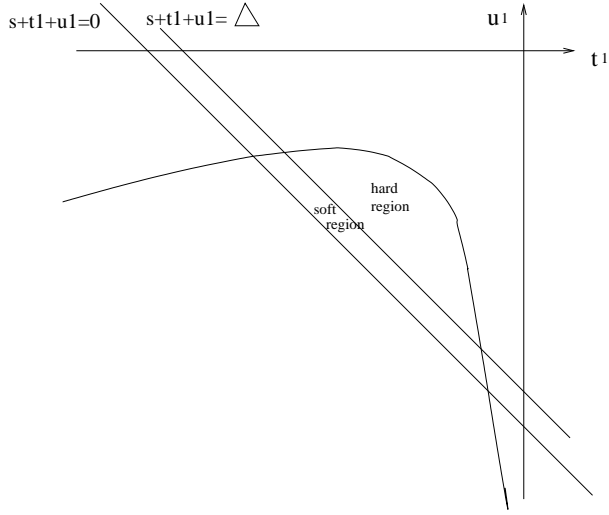}}
\caption{Dalitz plot for $\Phi+g\rightarrow Q+\bar{Q}+g$.}
\label{fig:Dalitz2}
\end{figure}

The Dalitz plot for $\Phi+g \rightarrow Q +\bar{Q}+ g$ is the same
as $\Phi+q \rightarrow Q +\bar{Q}+q$. But it is separated as soft
gluon emitting region and hard gluon emitting region as shown in
Fig.~(\ref{fig:Dalitz2}). The boundary line is $s_4=\Delta$ which
is an arbitrarily small value. That is, if $s_4$ is smaller
(larger) than $\Delta$, it corresponds to the region where soft
(hard) gluon are emitted.   Because all infrared divergences exit
in soft gluon emitting region, hard gluon region has only
collinear divergence. This collinear divergence is eliminated by
mass factorization.

\begin{eqnarray}
\acute{s}^2 \frac{d^2 \hat{\sigma}_{NLO-g}}{d t_1 du_1}&=&s^2
\frac{d^2 \sigma_{NLO-g}}{d t_1
du_1}\nonumber\\
&-&\frac{\alpha_s}{2 \pi} \int^1_0 \frac{dx}{x} P_{gg}(x) \bigg(
\frac{2}{D-4}+\gamma_E+\ln\frac{Q^2}{4\pi\mu^2} \bigg)
\hat{\acute{s}}^2 \frac{d^2
\hat{\sigma}_{LO}}{d\hat{t_1}du_1},\label{gmf}
\end{eqnarray}
where $P_{gg}(x)$ is a gluon to gluon splitting function, and may
be separated into hard part and soft part, which are proportional
to $\theta$ function and $\delta$ function respectively
\cite{Beenakker}.

\begin{eqnarray}
P_{gg}(x)&=& N_c  \theta(1-x-\delta) \bigg(
\frac{2}{1-x}+\frac{2}{x}-4+2x-2x^2 \bigg)\nonumber\\
&&\ \ \ \ +\ N_c \delta(1-x)\bigg(2\ln\delta+\frac{11}{6}-\frac{N_f}{3N_c}\bigg) \nonumber\\
&\equiv& \theta(1-x-\delta)P_{gg}^H
+\delta(1-x)P_{gg}^S.\label{mfs}
\end{eqnarray}

We ignore the factor proportional to $N_f$(the number of light
quark flavor) in the soft part, because it is suppressed in the
large $N_c$ limit. The boundary of hard and soft mass
factorization is $x=1-\delta$. $\delta$ is related to $\Delta$ by
$\delta=\Delta/(\acute{s}+t_1)$, because $x\leq 1-\delta$ means
$s_4 \geq \delta ( \acute{s}+t_1)$. After hard part mass
factorization, Eq.~({\ref{hard3}) becomes

\begin{eqnarray}
&&\frac{2^{9}}{3}\frac{1}{(4\pi)^2} g^4 m^2 (2 m_\Phi )^4
\bigg(\frac{\epsilon_o
}{m}\bigg)^{\frac{5}{2}}\frac{1}{\acute{s}+t_1}\nonumber\\
&&\ \ \ \ \ \ \ \ \ \ \times \bigg[\
\frac{\acute{s}-2m_\Phi\epsilon_o}{\acute{s}^3}\frac{1}{(1-X)\acute{s}}\bigg(\ln
\frac{s_4^2}{Q^2 (s_4+m^2)}+J^{(1,1)*}_0 \bigg) \nonumber\\
&&\ \ \ \ \ \ \ \ \ \ \ \
+\frac{\acute{s}-2m_\Phi\epsilon_o}{\acute{s}^3}\frac{1}{\acute{s}X}\bigg(\ln
\frac{s_4^2}{Q^2 (s_4+m^2)}+I^{(1,1)*}_0 \bigg) \nonumber\\
&&\ \ \ \ \ \ \ \ \ \ \ \ -\frac{2m_\Phi \epsilon_o}{\acute{s}^2}
\frac{1}{(\acute{s}X)^2}\bigg(\ln
\frac{s_4^2}{Q^2 (s_4+m^2)}+I^{(1,2)*}_0 \bigg) \nonumber\\
&&\ \ \ \ \ \ \ \ \ \ \ \
+2\frac{1}{(\acute{s}X)^3}\bigg(\ln\frac{s_4^2}{Q^2 (s_4+m^2)}+I^{(1,3)*}_0 \bigg) \nonumber\\
&&\ \ \ \ \ \ \ \ \ \ \ \ -(\acute{s}+4m_\Phi \epsilon_o
)\frac{1}{(\acute{s}X)^4}\bigg(\ln
\frac{s_4^2}{Q^2 (s_4+m^2)}+I^{(1,4)*}_0 \bigg) \nonumber\\
&&\ \ \ \ \ \ \ \ \ \ \ \
+\acute{s}(\acute{s}+2m_\Phi\epsilon_o)\frac{1}{(\acute{s}X)^5}\bigg(\ln
\frac{s_4^2}{Q^2 (s_4+m^2)}+I^{(1,5)*}_0 \bigg)
\nonumber\\
&&\ \ \ \ \ \ \ \ \ \ \ \ -2m_\Phi\epsilon_o \acute{s}^2
\frac{1}{(\acute{s}X)^6}\bigg(\ln \frac{s_4^2}{Q^2
(s_4+m^2)}+I^{(1,6)*}_0 \bigg)\ \bigg],\label{hardpart}
\end{eqnarray}
where $J^{(1,1)*}_0 \equiv a(\acute{A}+\acute{B})J^{(1,i)}_0/\pi =
a(X-1)J^{(1,i)}_0/\pi$. After mass factorization, collinear
divergence $1/(D-4)$ of Eq.~(\ref{hard3}) is removed.

\section{Soft part of the $\Phi + g \rightarrow Q + \bar{Q}+ g$ process}

\subsection{Differential cross section for soft gluon emitted part}

In hard gluon emitted region, differential cross section is the
sum of Eq.~(\ref{hard1}), Eq.~(\ref{hard2}), and
Eq.~(\ref{hardpart}). But in soft gluon emitted region,
$\acute{t}\rightarrow 0$, $\acute{u}\rightarrow 0$, and
$s_4\rightarrow 0$. In this limit, only soft and soft-collinear
divergent terms contribute and the differential cross section
becomes

\begin{eqnarray}
\acute{s}^2 \frac{d^2 \sigma^S}{dt_1 du_1}&\equiv&
\delta(s+t_1+u_1) \int^\Delta_0 ds_4 \acute{s}^2 \frac{d^2
\sigma}{dt_1
du_1}\nonumber\\
&=&\frac{2^9}{3}\frac{1}{(4\pi)^2}g^4 m^2 (2m_\Phi)^4
\bigg(\frac{\epsilon_o}{m}\bigg)^{\frac{5}{2}}\frac{\acute{s}-2m_\Phi
\epsilon_o}{\acute{s}^4} \delta(s+t_1+u_1)\nonumber\\
&&\times \bigg[ -\frac{1}{D-4}-\gamma_E -\ln
\frac{\Delta\sqrt{\acute{s}u_1 t_1-m_\Phi^2
t_1^2-m^2\acute{s}^2}}{4 \pi \mu^2 m\acute{s}}-\frac{1}{2}
\nonumber\\
&& +\frac{2}{(D-4)^2}+\frac{2}{D-4}\bigg(\gamma_E
+\ln\frac{\Delta\sqrt{\acute{s}u_1 t_1-m_\Phi^2
t_1^2-m^2\acute{s}^2}}{4 \pi \mu^2 m_\Phi
(\acute{s}+t_1)}+\frac{1}{2}\
\bigg)\nonumber\\
&&\ \ \ \ \ +\ln^2 \frac{\Delta\sqrt{\acute{s}u_1 t_1-m_\Phi^2
t_1^2-m^2\acute{s}^2}}{4 \pi m\mu^2 \acute{s}} \nonumber\\
&&\ \ \ \ \  +2\ln \frac{\Delta\sqrt{\acute{s}u_1 t_1-m_\Phi^2
t_1^2-m^2\acute{s}^2}}{4 \pi m\mu^2
\acute{s}}\cdot\ln\frac{m\acute{s}}{(\acute{s}+t_1)m_\Phi}\nonumber\\
&&\ \ \ \ \  +(2\gamma_E+1) \ln\frac{\Delta\sqrt{\acute{s}u_1
t_1-m_\Phi^2 t_1^2-m^2\acute{s}^2}}{4 \pi \mu^2 m_\Phi
(\acute{s}+t_1)}\nonumber\\
&&\ \ \ \ \  +\gamma_E^2+\gamma_E
-\frac{\pi^2}{6}+\frac{\Theta}{2}\ \bigg], \label{soft-g}
\end{eqnarray}
where

\begin{eqnarray}
\Theta &\equiv& 2 Li_2
\bigg(-\frac{\acute{B}+\sqrt{\acute{B}^2+\acute{C}^2}}{\acute{A}-\sqrt{\acute{B}^2+\acute{C}^2}}
\bigg) -2 Li_2 \bigg(
\frac{\acute{B}-\sqrt{\acute{B}^2+\acute{C}^2}}{\acute{A}+\acute{B}}\bigg)\nonumber\\
&&+\ln^2
\frac{\acute{A}-\sqrt{\acute{B}^2+\acute{C}^2}}{\acute{A}+\acute{B}}-\frac{1}{2}\ln^2
\frac{\acute{A}+\sqrt{\acute{B}^2+\acute{C}^2}}{\acute{A}-\sqrt{\acute{B}^2+\acute{C}^2}},
\end{eqnarray}
and we used the limiting values
$\acute{A}^2-\acute{B}^2-\acute{C}^2 \rightarrow s_4^2
m_\Phi^2/m^2$, $a(\acute{A}+\acute{B})\rightarrow \acute{s}
s_4^2/(2m^2)$, and
$(\acute{A}+\acute{B})^2/(\acute{A}^2-\acute{B}^2-\acute{C}^2)\rightarrow
[\acute{s} m/m_\Phi (\acute{s}+t_1)]^2$. For the definition of
soft gluon differential cross section, refer to \cite{Beenakker}.
In Eq.~(\ref{soft-g}), the first line in the square bracket comes
from soft divergence, and the others come from soft-collinear
divergence.  These soft divergences may be eliminated by adding
the mixed term of Born diagram and its one loop corrections.

\subsection{Effective four point vertex}

Before considering the one loop correction, it is helpful to
introduce the effective four point vertex. This vertex is attached
to a quarkonium, a gluon, a heavy quark, and a heavy antiquark
line and defined as

\begin{eqnarray}
M_{\mu \nu}^{(a)}(k)&\equiv& -g \sqrt{\frac{m_\Phi}{N_c}} \bigg[\
\vec{k}\cdot \frac{\partial \psi(p)}{\partial \vec{p}} g^{\nu
0}+\bigg(\frac{|\vec{p}|^2}{m}+\epsilon_0 \bigg) \frac{\partial
\psi(p)}{\partial p_j}g^{\nu j}\ \bigg]\nonumber\\
&&\times \frac{1+\gamma_0}{2}\gamma^i g_i^\mu
\frac{1-\gamma_0}{2}T^a.
\end{eqnarray}

It is just the leading order invariant amplitude given in
Eq.~(\ref{lo}), except that $k_{10}$ is substituted by
$|\vec{p}|^2/m+\epsilon_0$. Although
$|\vec{p}|^2/m+\epsilon_0=k_{10}$ to LO, it is not so in general.
Using this effective vertex, the matrix element for the process
$\Phi + g \rightarrow Q + \bar{Q}+ g$ represented in
Fig.~(\ref{fig:nlog}) can be reproduced by five diagrams shown in
Fig.~(\ref{fig:newvertex}). Specifically,

\begin{figure}
\centerline{\includegraphics[width=12cm]{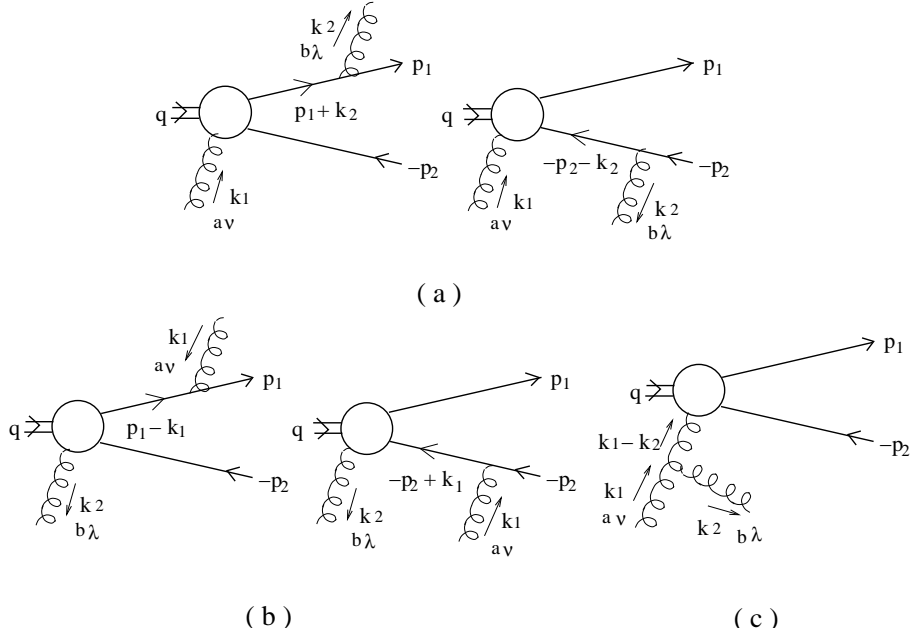}}
\caption{Diagrams for $\Phi+g \rightarrow Q+\bar{Q}+g$ using
four-point vertex. (a), (b), and (c) are diagrams for $M_1$,
$M_2$, and $M_3$ respectively.} \label{fig:newvertex}
\end{figure}

\begin{eqnarray}
M_1^{\mu \nu \lambda (a,b)} &=& -g \bar{u}(p_1)\bigg[\
\gamma^\lambda T^b
\Delta (p_1+k_2)M^{\mu \nu (a)}(k_1) \nonumber\\
&&\ \ \ \ \ \ +M^{\mu \nu (a)}(k_1)\Delta(-p_2-k_2) \gamma^\lambda
T^b \ \bigg]v(p_2) \nonumber\\
&=& g \frac{g^{\lambda 0}}{k_{20}+i\varepsilon} \bar{u}(p_1)
M^{\mu \nu}(k_1)[T^a,T^b]v(p_2),
\end{eqnarray}
where $M^{\mu \nu (a)}(k_1)=M^{\mu \nu }(k_1)T^a$.

$M_2$ is the same as $M_1$ with $(k_1, a, \nu)$ and $(-k_2, b,
\lambda)$ exchanged.

\begin{eqnarray}
M_2^{\mu \nu \lambda (a,b)}=g \frac{g^{\nu
0}}{-k_{10}+i\varepsilon} \bar{u}(p_1) M^{\mu
\lambda}(-k_2)[T^b,T^a]v(p_2).
\end{eqnarray}

$M_3$ is a diagram which emits a gluon from the external gluon
leg.

\begin{eqnarray}
M_3^{\mu \nu \lambda (a,b)}&=&-i g f^{abc} \bar{u}(p_1)M^{\mu
\sigma
(c)}(k_1-k_2)v(p_2)\frac{1}{(k_1-k_2)^2+i\varepsilon}\nonumber\\
&&\ \ \ \ \ \ \times [(k_1+k_2)_\sigma g^{\nu
\lambda}+(k_1-2k_2)^\nu g^\lambda_\sigma +(-2k_1+k_2)^\lambda
g^\nu_\sigma ]
\end{eqnarray}

The sum of all diagrams is exactly the same as Eq.~(\ref{amp-g}).

\begin{eqnarray}
M_1^{\mu \nu \lambda (a,b)
}+M_2^{\mu \nu \lambda (a,b)}+M_3^{\mu \nu \lambda (a,b)}=M^{\mu \nu \lambda (a,b) }_{NLO-g}
\end{eqnarray}

Introduction of this effective vertex has some benefits.  It makes the calculation
much easier and one does not need to consider the inner structure
of the four point vertex, which is very complicated when
considering the one loop corrections.

\subsection{One loop correction}

\begin{figure}
\centerline{\includegraphics[width=12cm]{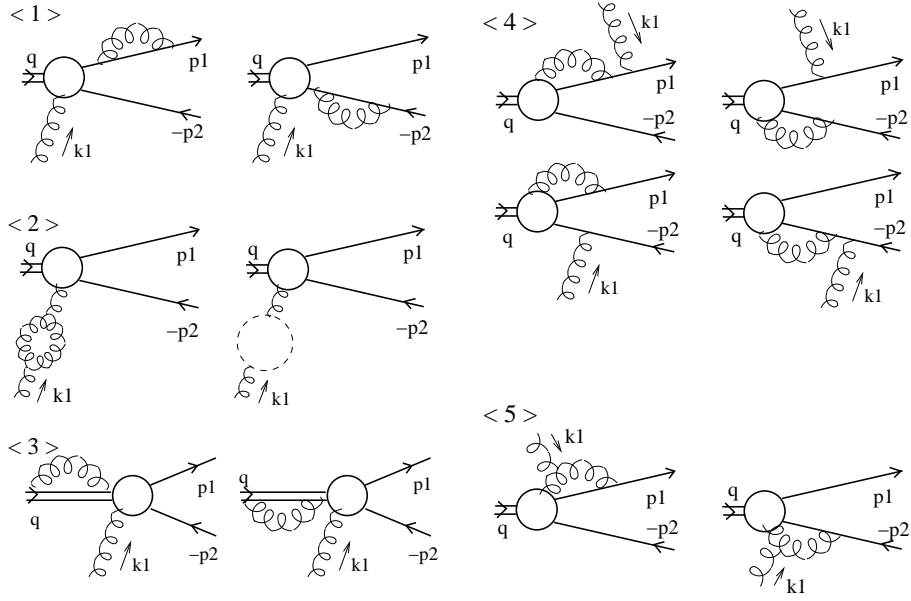}} \caption{One
loop corrections.}\label{fig:bv}
\end{figure}

Several comments are in order before considering the one loop
correction.  In dimensional regularization,

\begin{eqnarray}
\frac{\Gamma(D/2)}{\pi^{D/2}i}\int \frac{d^D
q}{(-q)^\alpha}=\frac{\Lambda^{D_I-2\alpha}}{D_I/2-\alpha}
-\frac{\Lambda^{D_U-2\alpha}}{D_U/2-\alpha}.
\end{eqnarray}
$D_I$ is the dimension which regularizes the infrared divergence,
and $D_U$ the ultraviolet divergence.  $\Lambda$ is the cutoff of
the momentum integral.  Generally this integration is zero. But we
have left over $\alpha=2$ case, because it shows clearly the
cancellation of infrared divergence and ultraviolet divergence
separately.  Second, in contrast to the order of typical loop
momentum  appearing in Eq.~(\ref{hardg}), which is  of $O(mg^2)$,
we set the order of gluon loop momentum in the one loop
corrections to be of $O(mg^4)$.   This is to explicitly separate
the soft part that cancels the soft divergence coming from emitted
gluons of $O(mg^4)$. Keeping these comments, all divergences may
be eliminated systematically.

Diagram $<1>$ of Fig.~(\ref{fig:bv}) is the one loop correction of
heavy quark and antiquark external lines. The product of on-shell
heavy quark propagator and its self energy is

\begin{eqnarray}
i\Delta(p_1)\Sigma(p_1)&=&-ig^2\int\frac{d^D
k}{(2\pi)^D}\gamma^\mu \Delta(p_1+k)\gamma_\mu \Delta(p_1)T^a T^a
\frac{1}{k^2+i\varepsilon}\nonumber\\
&=&-ig^2 T^a T^a \frac{1+\gamma_0}{2}\int\frac{d^D k}{(2\pi)^D}
\frac{4m^2}{(p_1+k)^2-m^2+i\varepsilon}\frac{1}{p_1^2-m^2+i\varepsilon}\frac{1}{k^2+i\varepsilon}.
\end{eqnarray}

We assumed that $p_1$ is slightly off-shell, and used the  heavy
quark propagators as in Eq~(\ref{qpropa}).  Moreover,
$\Delta(p_1+k)$ is replaced by $[\Delta(p_1+k)+\Delta(p_1-k)]/2$.
Then one has,

\begin{eqnarray}
i\Delta(p_1)\Sigma(p_1)&=& ig^2 T^a T^a
\frac{1+\gamma_0}{2}\int\frac{d^D
k}{(2\pi)^D}\frac{1}{(k_0+i\varepsilon)(k_0-i\varepsilon)(k^2+i\varepsilon)}\nonumber\\
&=&g^2 \frac{1}{(4\pi)^{\frac{D-1}{2}}\Gamma(\frac{D-1}{2})}
\bigg[\frac{\Lambda^{D-4}}{D_I-4}-\frac{\Lambda^{D-4}}{D_U-4}\bigg]
T^a T^a \frac{1+\gamma_0}{2}.  \label{self1}
\end{eqnarray}

In the $k_0$ contour integration, the residue at
$k_0=i\varepsilon$ or at $k_0=-i\varepsilon$ makes the one loop
correction pure imaginary. Therefore, in these cases, the mixed
term of LO and its one loop correction vanishes. The self energy
of the antiquark is the same except that $(1+\gamma_0)/2$ is
replaced by $(1-\gamma_0)/2$. However it has the same contribution
to the differential cross section, because spinor wavefunction
$u(p_1)$ is proportional to $(1+\gamma_0)/2$, while $v(p_2)$ is to
$(1-\gamma_0)/2$ in the heavy quark limit.

As can be seen from Eq.~(\ref{self1}), the renormalization
constant of the heavy quark mass has no divergence. Only the
renormalization constants of the heavy quark and antiquark fields
have both ultraviolet and infrared divergences. The differential
cross section from the mixed term of Born diagram and the same
diagram but with the self energy insertion to the  external heavy
quark or antiquark line is

\begin{eqnarray}
\acute{s}^2 \frac{d^2 \sigma^{BV1}}{dt_1 du_1}&=& \frac{2^9}{3}
\frac{1}{(4 \pi)^2} g^4 m^2 (2m_\Phi)^4 \bigg(
\frac{\epsilon_o}{m}\bigg)^{\frac{5}{2}} \frac{\acute{s}-2m_\Phi
\epsilon_o}{\acute{s}^4} \nonumber\\
&&
\times\delta(s+t_1+u_1)\bigg[\frac{1}{D_I-4}-\frac{1}{D_U-4}\bigg].
\label{self}
\end{eqnarray}

Note that the result was divided by $2$, because half of the
divergence is used for the renormalization constant of external
quark antiquark wavefunction.  It may be checked that the above
infrared divergence cancels the soft divergence of
Eq.~(\ref{soft-g}).

Diagram $<2>$ of Fig.~(\ref{fig:bv}) is the external gluon line
correction. The self energy of the external gluon is

\begin{eqnarray}
\Pi_{\alpha \nu}^{ba}(k_1)&=& \frac{1}{2}g^2 N_c \delta^{ba}\int
\frac{d^D k}{(2\pi)^D}\bigg[ (2k^2+2k\cdot k_1+5k_1^2)g_{\alpha
\nu} +(4D-8)k_\alpha k_\nu \nonumber\\
&& +(2D-4)(k_\alpha k_{1\nu}+k_\nu
k_{1\alpha})+(D-6)k_{1\alpha}k_{1\nu}\bigg]\frac{1}{(k+k_1)^2
k^2}\nonumber\\
&=&\frac{1}{2}g^2 N_c \delta^{ba}g_{\alpha \nu}\int_0^1 dx \int
\frac{d^D
k}{(2\pi)^D}\bigg[\frac{(6-8/D)k^2}{(k^2-x(1-x)(-k_1^2))^2}\nonumber\\
&&\ \ \ \ \ \ \ \ \ \ \ \ \ \ \ \ \ \ \ \ \ \ \ \ \ \
+\frac{(2x^2-2x+5)k_1^2}{(k^2-x(1-x)(-k_1^2))^2}\bigg].
\end{eqnarray}

Here, we ignored quark loop contribution, as it is suppressed in
the large $N_c$ limit.  Terms proportional to $k_{1\alpha}$ and
$k_{1 \nu}$ will vanish due to the current conservation condition
of the LO amplitude.   Assuming $k_1$ to be slightly off-shell,
the gluon self energy may be expanded with respect to $k_1^2$.

\begin{eqnarray}
\frac{k^2}{(k^2-x(1-x)(-k_1^2))^2}=\frac{1}{k^2}+\frac{2x(1-x)(-k_1^2)}{k^4}+\cdot\cdot\cdot, \nonumber\\
\frac{1}{(k^2-x(1-x)(-k_1^2))^2}=\frac{1}{k^4}+\frac{2x(1-x)(-k_1^2)}{k^6}+\cdot\cdot\cdot.
\end{eqnarray}

Keeping only the $1/k^4$ terms,

\begin{eqnarray}
\frac{-i}{k_1^2}\Pi_{\alpha \nu}^{ba}(k_1)=\delta^{ba}g_{\alpha
\nu}\frac{g^2 N_c}{(4\pi)^{D/2}\Gamma(D/2)}\frac{10}{3}
\bigg[\frac{\Lambda^{D-4}}{D_I-4}-\frac{\Lambda^{D-4}}{D_U-4}\bigg].
\label{self2}
\end{eqnarray}

The differential cross section obtained from multiplying this and
the Born cross section is

\begin{eqnarray}
\acute{s}^2 \frac{d^2 \sigma^{BV2}}{dt_1 du_1}&=& \frac{2^9}{3}
\frac{1}{(4 \pi)^2} g^4 m^2 (2m_\Phi)^4 \bigg(
\frac{\epsilon_o}{m}\bigg)^{\frac{5}{2}} \frac{\acute{s}-2m_\Phi
\epsilon_o}{\acute{s}^4} \nonumber\\
&&
\times\delta(s+t_1+u_1)\frac{5}{6}\bigg[\frac{1}{D_I-4}-\frac{1}{D_U-4}\bigg].\label{bv2}
\end{eqnarray}

Again, the result was divided by $2$ for the same reason.

Diagram $<3>$ of Fig.~(\ref{fig:bv}) is the quarkonium external
line correction.  For quarkonium external line, there is no direct
one loop correction in QCD. But we assume that its one loop
correction is the same as that of heavy quark and antiquark lines
in Eq.~(\ref{self}).  This assumption can be proven to be true
from noting that the quarkonium is the bound state of a heavy
quark and an antiquark, and therefore its field operator is a
composite operator composed of a quark and an antiquark field.

Diagram $<4>$ of Fig.~(\ref{fig:bv}) is another type of one loop
correction. However it vanishes and has no contribution.

\begin{eqnarray}
M_{\mu \nu}^{(a) V4}&=&g^2 C_F \int \frac{d^D
k}{(2\pi)^D}\frac{-i}{k^2+i\varepsilon}\nonumber\\
&\times& \bar{u}(p_1)\bigg[ \gamma^\nu \Delta(p_1-k_1)
\gamma^\beta
\Delta(p_1-k_1+k)M_{\beta \mu}(k)\nonumber\\
&&\ \ \ \ \ \ \ \  +\gamma^\nu \Delta(p_1-k_1) M_{\beta
\mu}(k)\Delta(-p_2-k)\gamma^\beta \nonumber\\
&&\ \ \ \ \ \ \ \  +\gamma^\beta \Delta(p_1+k)M_{\beta
\mu}(k)\Delta(-p_2+k_1)\gamma^\nu \nonumber\\
&&\ \ \ \ \ \ \ \  +M_{\beta \mu}(k)\Delta(-p_2+k_1-k)\gamma^\beta
\Delta(-p_2+k_1)\gamma^\nu \bigg]T^a v(p_2)\nonumber\\
&=& g^3 C_F \sqrt{\frac{m_\Phi}{N_c}}\bar{u}(p_1)
\frac{1+\gamma_0}{2}\gamma^i g_i^\mu
\frac{1-\gamma_0}{2}T^a v(p_2)\nonumber\\
&&\times \int \frac{d^{D-1} k}{(2\pi)^{D-1}}\vec{k}\cdot
\frac{\partial \psi (p)}{\partial
\vec{p}}\frac{1}{k_{10}}\bigg(\frac{1}{|\vec{k}|^2}-\frac{1}{k_{10}|\vec{k}|}+
\frac{1}{k_{10}(|\vec{k}|+k_{10})}\bigg)= 0.
\end{eqnarray}

Finally diagram $<5>$ of Fig.~(\ref{fig:bv}) is the vertex
correction. Soft collinear divergence of Eq.~(\ref{soft-g}) is
eliminated by these diagrams.  Such cancellation can be
anticipated, because the soft collinear divergence of
Eq.~(\ref{soft-g}) and the cross section coming from multiplying
diagram $<5>$ with a Born diagram are coming from the different
cutting of a two loop diagram for the self energy of the
quarkonium with no infrared divergence.   The amplitude becomes,

\begin{eqnarray}
M_{\mu \nu}^{(a) V5} &=& g^2 f^{abc}\int \frac{d^D
k}{(2\pi)^D}\bar{u}(p_1)\bigg[ M_{\mu
\alpha}^{(c)}(k+k_1)\Delta (-p_2-k)\gamma_\beta T^b \nonumber\\
&&\ \ \ \ \ \ \ \ \ \ \ \ \ \ \ \ \ \ \ \ + \gamma_\beta T^b
\Delta (p_1+k)M_{\mu
\alpha}^{(c)}(k+k_1)\bigg]v(p_2)\nonumber\\
&& \times \frac{(-2k_1-k)^\beta g^{\alpha}_{\nu} +(k_1-k)^\alpha
g_{\nu}^{\beta}+(k_1+2k)_\nu g^{\alpha
\beta}}{(k^2+i\varepsilon)((k_1+k)^2+i\varepsilon)}\nonumber\\
&=& g^3 \sqrt{m_\Phi N_c}\bar{u}(p_1) \frac{1+\gamma_0}{2}\gamma^i
g_i^\mu \frac{1-\gamma_0}{2}T^a v(p_2)\int \frac{d^{D-1}
k}{(2\pi)^{D-1}}\nonumber\\
&& \times \bigg[\frac{2k_{10}\frac{\partial \psi(p)}{\partial
\vec{p}}\cdot \vec{k_1}g_{\nu
0}+k_{10}(2k_{10}-|\vec{k}|)\frac{\partial \psi(p)}{\partial
p_j}g_\nu^j-2k_j \frac{\partial \psi(p)}{\partial \vec{p}}\cdot
(\vec{k}+\vec{k_1})g_\nu^j}{4|\vec{k}|^2(k_{10}|\vec{k}|+\vec{k}\cdot\vec{k_1})}\nonumber\\
&&\ \ \ \ \ +\frac{-2k_{10}\frac{\partial \psi(p)}{\partial
\vec{p}}\cdot \vec{k_1}g_{\nu
0}-k_{10}(k_{10}-|\vec{k}|)\frac{\partial \psi(p)}{\partial
p_j}g_\nu^j+2(k-k_1)_j \frac{\partial \psi(p)}{\partial
\vec{p}}\cdot \vec{k}
g_\nu^j}{4|\vec{k}|(k_{10}+|\vec{k}|)(k_{10}|\vec{k}|+\vec{k}\cdot\vec{k_1})}\bigg].
\end{eqnarray}

Two terms of the last equation are residues at
$k_0=-|\vec{k}|+i\varepsilon$, and at
$k_0=-k_{10}-|\vec{k}+\vec{k_1}|+i\varepsilon$ respectively. In
the second term, $\vec{k}+\vec{k_1}$ was substituted by $\vec{k}$.
If each momentum is set to be

\begin{eqnarray}
q&=&(m_\Phi, 0, ... , 0, 0, 0, 0)\nonumber\\
k_1&=&(k_{10}, 0, ... , 0, 0, 0,k_{10}) \nonumber\\
p_1&=&(E_1, 0, ... , 0, 0, |\vec{p}|\sin\theta_1,
|\vec{p}|\cos\theta_1)\nonumber\\
p_2&=&(E_2, 0, ... , 0, 0, -|\vec{p}|\sin\theta_1,
-|\vec{p}|\cos\theta_1+k_{10})\nonumber\\
k&=&(k_{0}, 0, ... ,
0,|\vec{k}|\sin\theta\sin\varphi,|\vec{k}|\sin\theta\cos\varphi,|\vec{k}|\cos\theta),
\end{eqnarray}
then,
\begin{eqnarray}
M_{\mu \nu}^{(a) V5} &=&g^3 \sqrt{m_\Phi N_c}\bar{u}(p_1)
\frac{1+\gamma_0}{2}\gamma^i g_i^\mu \frac{1-\gamma_0}{2}T^a
v(p_2)\int \frac{d^{D-1}
k}{(2\pi)^{D-1}}\nonumber\\
&&
\times\bigg[\frac{k_{10}^2}{2|\vec{k}|^2(k_{10}+|\vec{k}|)(|\vec{k}|k_{10}+\vec{k}\cdot\vec{k_1})}
\bigg( \vec{k_1}\cdot \frac{
\partial \psi (p)} {\partial \vec{p}}
g_{\nu 0} + k_{10} \frac{\partial \psi(p)}{\partial p_j } g^j_\nu
\bigg)\nonumber\\
&&\ \ \ \ \ \ +g_{\nu 2}\frac{1
}{2k_{10}(k_{10}+|\vec{k}|)}|\vec{p}|\sin\theta_1 \frac{\partial
\psi(p)}{\partial \vec{p}^2}\bigg],\label{v5}
\end{eqnarray}
where $g_{\nu 2}$ means transverse direction with respect to
$\vec{k_1}$.  It is manifest that current conservation condition
$k_1^\nu M_{\mu \nu}^{(a) V}=0$ is satisfied from Eq.~(\ref{v5}),
because Eq.~(\ref{self1}) and Eq.~(\ref{self2}) are intrinsically
zero. The differential cross section from $M^{V5}$, after
multiplying it with the Born diagram, is

\begin{eqnarray}
\acute{s}^2 \frac{d^2 \sigma^{BV5}}{dt_1 du_1}&=&
\frac{2^9}{3}\frac{1}{(4\pi)^2}g^4 m^2 (2m_\Phi)^4
\bigg(\frac{\epsilon_o}{m}\bigg)^{\frac{5}{2}}\frac{\acute{s}-2m_\Phi
\epsilon_o}{\acute{s}^4}\delta(s+t_1+u_1)\nonumber\\
&&\times \bigg[-\frac{2}{(D-4)^2}-\frac{2}{D-4}\bigg(\gamma_E
+\ln\frac{\sqrt{\acute{s}u_1 t_1-m_\Phi^2 t_1^2-m^2\acute{s}^2}}{4
\pi \mu^2 m_\Phi }+\frac{1}{2}\ \bigg)\nonumber\\
&&-\ln^2 \frac{\sqrt{\acute{s}u_1 t_1-m_\Phi^2
t_1^2-m^2\acute{s}^2}}{4 \pi \mu^2 m_\Phi }-(2\gamma_E+1)\ln
\frac{\sqrt{\acute{s}u_1 t_1-m_\Phi^2
t_1^2-m^2\acute{s}^2}}{4 \pi \mu^2 m_\Phi }\nonumber\\
&&-\frac{\pi^2}{6}-\gamma_E^2-\gamma_E\nonumber\\
&&+\frac{1}{D_U-4}+\ln\frac{\sqrt{\acute{s}u_1 t_1-m_\Phi^2
t_1^2-m^2\acute{s}^2}}{4 \pi \mu^2 m_\Phi }+\gamma_E-\frac{1}{2}\
\ \bigg].\label{bv5}
\end{eqnarray}

The double pole $1/(D-4)^2$ is cancelled with that of
Eq.~(\ref{soft-g}).

\subsection{Coupling constant renormalization and soft part mass factorization}

The ultraviolet divergence in one loop correction may be removed
by renormalization of the coupling constant $g$.

\begin{eqnarray}
g_b \rightarrow g\bigg[ 1+\frac{\alpha_s}{8\pi}\bigg(
\frac{2}{D-4}+\gamma_E+\ln \frac{m^2}{4\pi
\mu^2}\bigg)\beta_0\bigg],
\end{eqnarray}
where the  renormalization scale is set to the heavy quark mass.
$\beta_0=\frac{11}{3}N_c$ in the large $N_c$ limit.

The sum of soft differential cross section Eq.~(\ref{soft-g}) and
the terms obtained by multiplying the Born term and its one loop
correction Eq.~(\ref{self}), Eq.~(\ref{bv2}), Eq.~(\ref{bv5})
still has collinear divergence, where Eq.~(\ref{self}) should be
doubled because of quarkonium external leg correction. This
remaining divergence is removed by the soft mass factorization in
Eq.~(\ref{mfs}), which corresponds to substituting the second part
of Eq.~(\ref{mfs}) proportional to $\delta(1-x)$ into
Eq.~(\ref{gmf}). becomes divergence-free.

\begin{eqnarray}
\acute{s}^2 \frac{d^2 \sigma^{S}}{dt_1 du_1}+\acute{s}^2 \frac{d^2
\sigma^{BV}}{dt_1 du_1} &=& \frac{2^9}{3}\frac{1}{(4\pi)^2}g^4 m^2
(2m_\Phi)^4
\bigg(\frac{\epsilon_o}{m}\bigg)^{\frac{5}{2}}\frac{\acute{s}-2m_\Phi
\epsilon_o}{\acute{s}^4}\delta(s+t_1+u_1)\nonumber\\
&&\times \bigg[\ \ln^2 \delta +2\ln \delta \ln
\frac{\acute{s}}{m_\Phi Q}-\ln \delta -\ln^2 \frac{m \acute{s}}{m_\Phi (\acute{s}+t_1)}\nonumber\\
&&\ \ \ +\ln \frac{m \acute{s}}{m_\Phi (\acute{s}+t_1)}
-\frac{\pi^2}{3}-1+\frac{\Theta}{2}+\frac{11}{6}\ln\frac{m}{Q}\ \
\bigg],  \label{softpart}
\end{eqnarray}
where $\frac{d^2\sigma^{BV}}{dt_1 du_1}$ is the sum of
$\frac{d^2\sigma^{BV1}}{dt_1 du_1}$ through
$\frac{d^2\sigma^{BV5}}{dt_1 du_1}$.

Below is the summary of the elementary total cross
section for $\Phi + g \rightarrow Q + \bar{Q} + g$.

\begin{eqnarray}
\sigma_{NLO-g}&=&\lim_{\Delta \rightarrow
0}\frac{1}{\acute{s}^2}\int_{-\frac{\acute{s}}{2s}(s-\Delta+\sqrt{(s-\Delta)^2-4m^2s})}^{-\frac{\acute{s}}{2s}(s-\Delta-\sqrt{(s-\Delta)^2-4m^2s})}dt_1
\int_{\Delta-s-t_1}^{\frac{m_\Phi^2 t_1^2
+m^2\acute{s}^2}{t_1\acute{s}}} du_1\ \  \acute{s}^2 \frac{d^2
\sigma^{H}}{dt_1
du_1}\nonumber\\
&+&\lim_{\Delta \rightarrow
0}\frac{1}{\acute{s}^2}\int_{-\frac{\acute{s}}{2}(1+\sqrt{1-4m^2/s})}^{-\frac{\acute{s}}{2}(1-\sqrt{1-4m^2/s})}dt_1
\int_{-s-t_1}^{\Delta-s-t_1}du_1 \ \bigg( \acute{s}^2 \frac{d^2
\sigma^{S}}{dt_1 du_1}+\acute{s}^2 \frac{d^2 \sigma^{BV}}{dt_1
du_1} \bigg).
\end{eqnarray}

Here, $\acute{s}^2 \frac{d^2 \sigma^{H}}{dt_1 du_1}$ is the sum of
Eq.~(\ref{hard1}), Eq.~(\ref{hard2}), Eq.~(\ref{hardpart}). The
first line and the second line depends on $\Delta$ (or $\delta$).
But their sum is independent of it, because it appears as the
lower cut in the first line and as the upper cut in the second
line.   The results are shown in Fig.~(\ref{fig:eleg}).

\section{Upsilon dissociation cross section}

\begin{figure}
\centerline{ \
\includegraphics[width=6cm, angle=270]{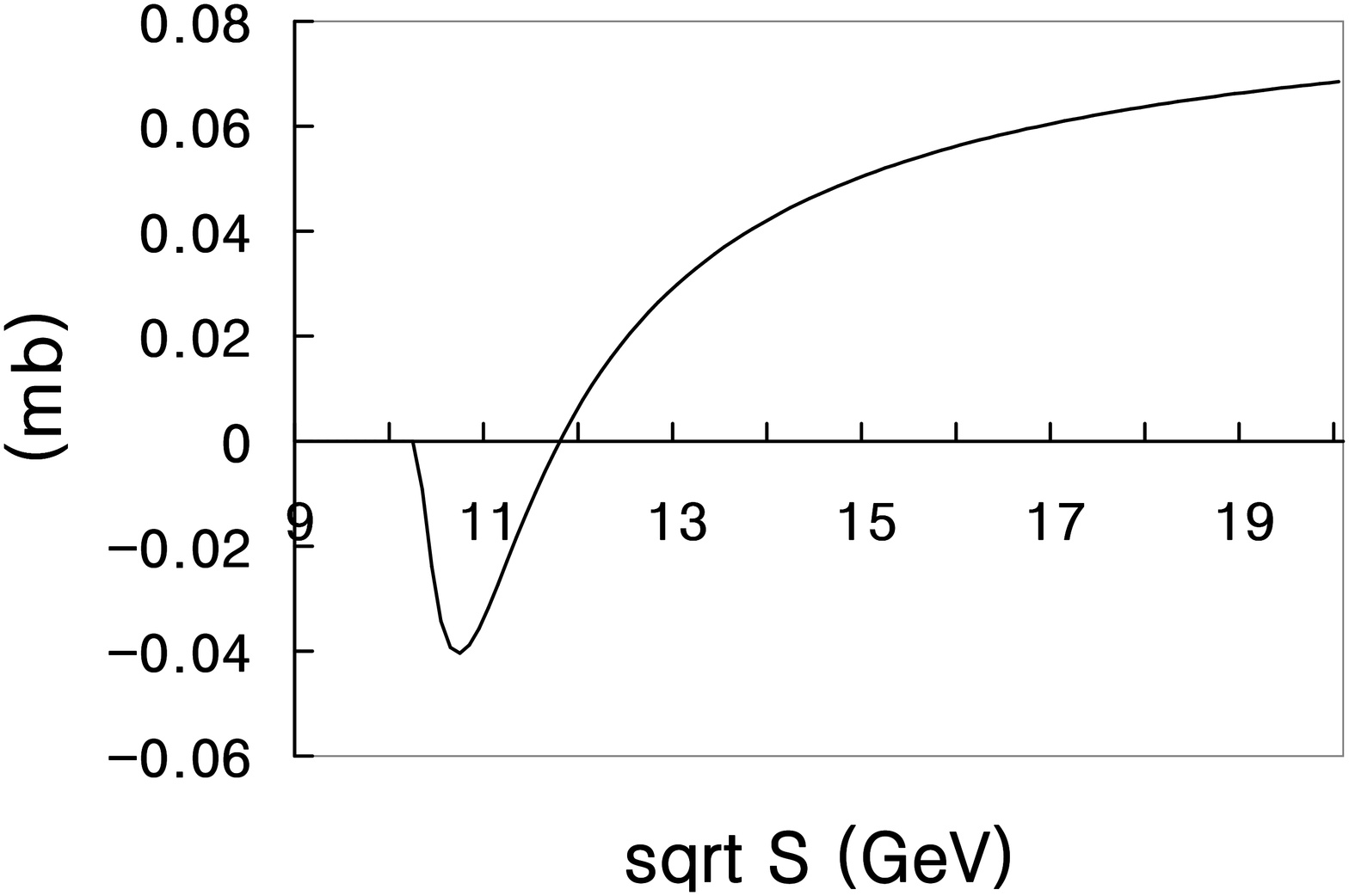}
\hfill
\includegraphics[width=6cm, angle=270]{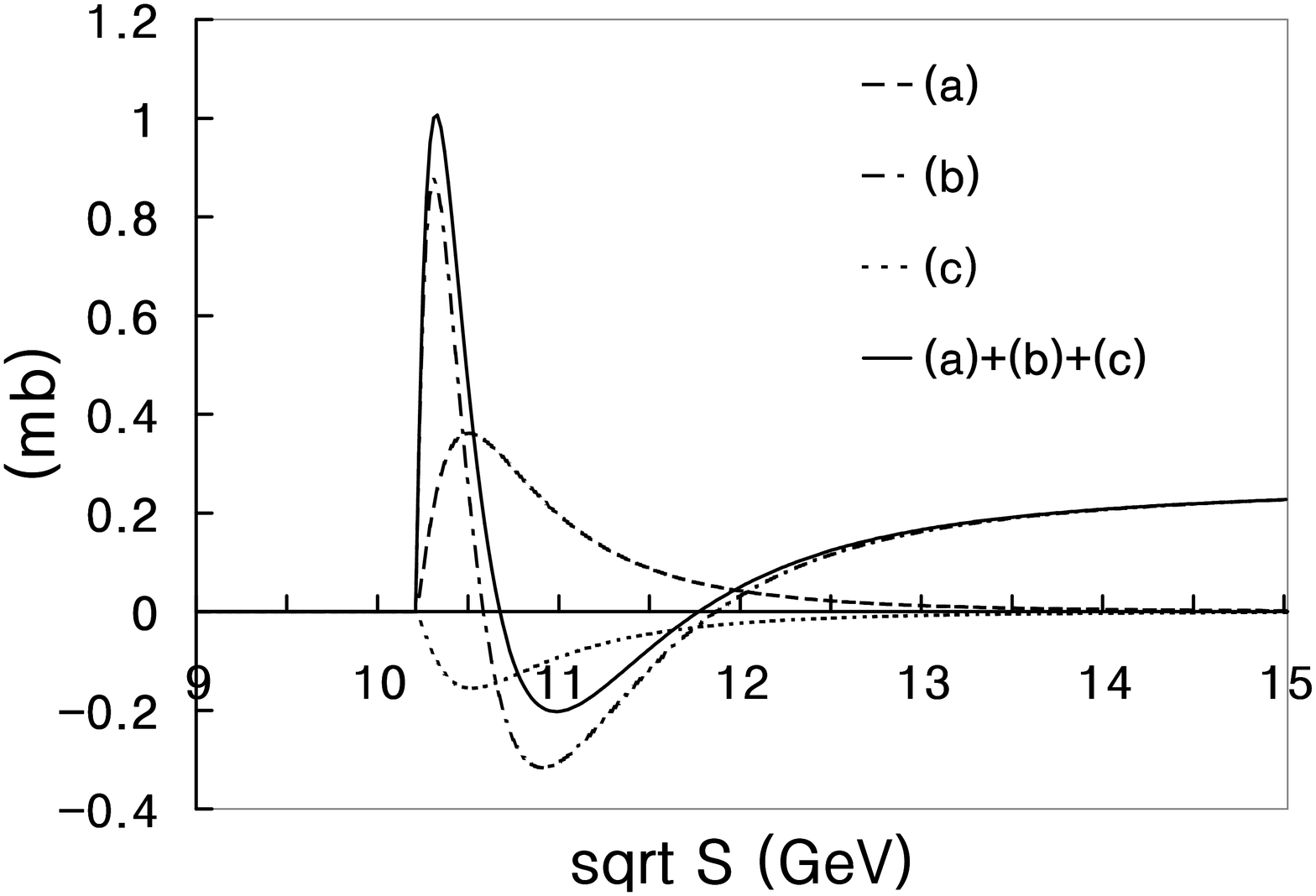}
}\caption{The left figure is the elementary cross section for
$\Phi+q \rightarrow Q+\bar{Q}+q$ and the right figure is that for
$\Phi+g \rightarrow Q+\bar{Q}+(g)$.  In the right figure, (a) the
dashed line is the Born term given in Eq.~(\ref{locs}). (b) The Dash
dotted line is the hard gluon part, namely, the integration of the
sum of Eq.~(\ref{hard1}), Eq.~(\ref{hard2}), and
Eq.~(\ref{hardpart}) over the hard part of the  Dalitz plot plus
the integration of the `$\ln \delta$' dependent part of
Eq.~(\ref{softpart}) over the soft part of the Dalitz plot.  (c) The
Dotted line is the soft plus one loop correction, namely, the
integration of Eq.~(\ref{softpart}) excluding the `$\ln \delta$'
dependent part over the soft part of the Dalitz plot.   The solid
line is the sum of (a), (b), and (c).} \label{fig:eleg}
\end{figure}
As an explicit application, the above result is applied to the
upsilon dissociation cross section. The two independent parameters
of the theory is determined by fitting the physical masses of
$m_{\Upsilon(1S)}$ and $m_{\Upsilon(2S)}$ to the energies of the
Coulomb bound states.  Specifically, from  the relation
$m_{\Upsilon(2S)}-m_{\Upsilon(1S)}=3/4 \epsilon_o$, the binding
energy is found to be $750  MeV$.   Also,  the bottom quark mass
is found to be $5.1\ GeV$ from equating it to
$(m_{\Upsilon(1S)}+\epsilon_o)/2$.  The coupling constant $g$ is
then found to be $2.53$ from $g^2=16\pi/N_c\sqrt{\epsilon_o/m}$.

The left and right graphs in Fig.~(\ref{fig:eleg}) represent the
elementary total cross sections of $\Phi+q \rightarrow Q +\bar{Q}
+q$ and $\Phi+g \rightarrow Q +\bar{Q} +g$ respectively.
In both graphs, there are regions of energy where the cross
sections become negative.   These negative cross sections
originate from mass factorization, where finite parts of the
differential cross section have been subtracted out and put into
the definition of the distribution function.   Therefore, the
cross section becomes physical only after folding the elementary
cross sections with the parton distribution functions (PDF) and
adding them to the LO contribution.

To obtain the total cross section, we used the MRST2001LO PDF
\cite{mrstlo} for the LO result, and  the MRST2001NLO PDF
\cite{mrstnlo} for the NLO.  We used the PDF calculated in the
$\overline{MS}$ scheme, because our perturbative calculations,
including the subtractions in the mass factorization, were
performed in the $\overline{MS}$ scheme.  If different schemes
were used in the perturbative calculation and in the PDF, the
scheme dependent finite pieces would not match, and the result
would be inconsistent.   In the original scheme of Peskin, the
scale of the PDF was be taken to be the binding energy of the
system, which is 0.75 GeV in the present system.   However, in the
present example, we will take it to be $1.25\ GeV^2$, which is the
minimum $Q^2$ scale in the MRST PDF.

The left graph of Fig.~(\ref{fig:result}) shows the total
dissociation cross section to LO and to NLO.  The ratio between
the cross sections calculated to  NLO and to LO, plotted in right
graph of Fig.~(\ref{fig:result}), shows that perturbative QCD
approach is acceptable only in a limited region of energy and
large corrections exist in the threshold region.

The separation scale of $1.25\ GeV^2$ is low, making it
questionable whether one can apply the present formalism to the
Upsilon system.   On the other hand, one can not take the
separation scale to be arbitrarily large in this example, as one
would invalidate all the counting schemes and the non relativistic
approximations used in deriving the formula.   Nevertheless, to
asses the uncertainties associated with takin the scale of the PDF
to be low, we modified the scale $Q^2$ of PDF to $2.0\ GeV^2$, and
compared the result to that obtained with $Q^2=1.25\ GeV^2$. As
shown in Fig.~(\ref{fig:modi}), the total cross section changes by
less than 10\% for $ \sqrt{s} < 25 GeV$.   As shown in the right
graph of Fig.~(\ref{fig:modi}), $ 25 GeV$ is also the upper limit
of the window of energy region where the ratio between NLO to LO
is minimal.  Hence, although the scale dependence is non
negligible in the present example, the uncertainties are within
the estimated errors coming from the perturbative expansion.

\begin{figure}
\centerline{
\includegraphics[width=6cm, angle=270]{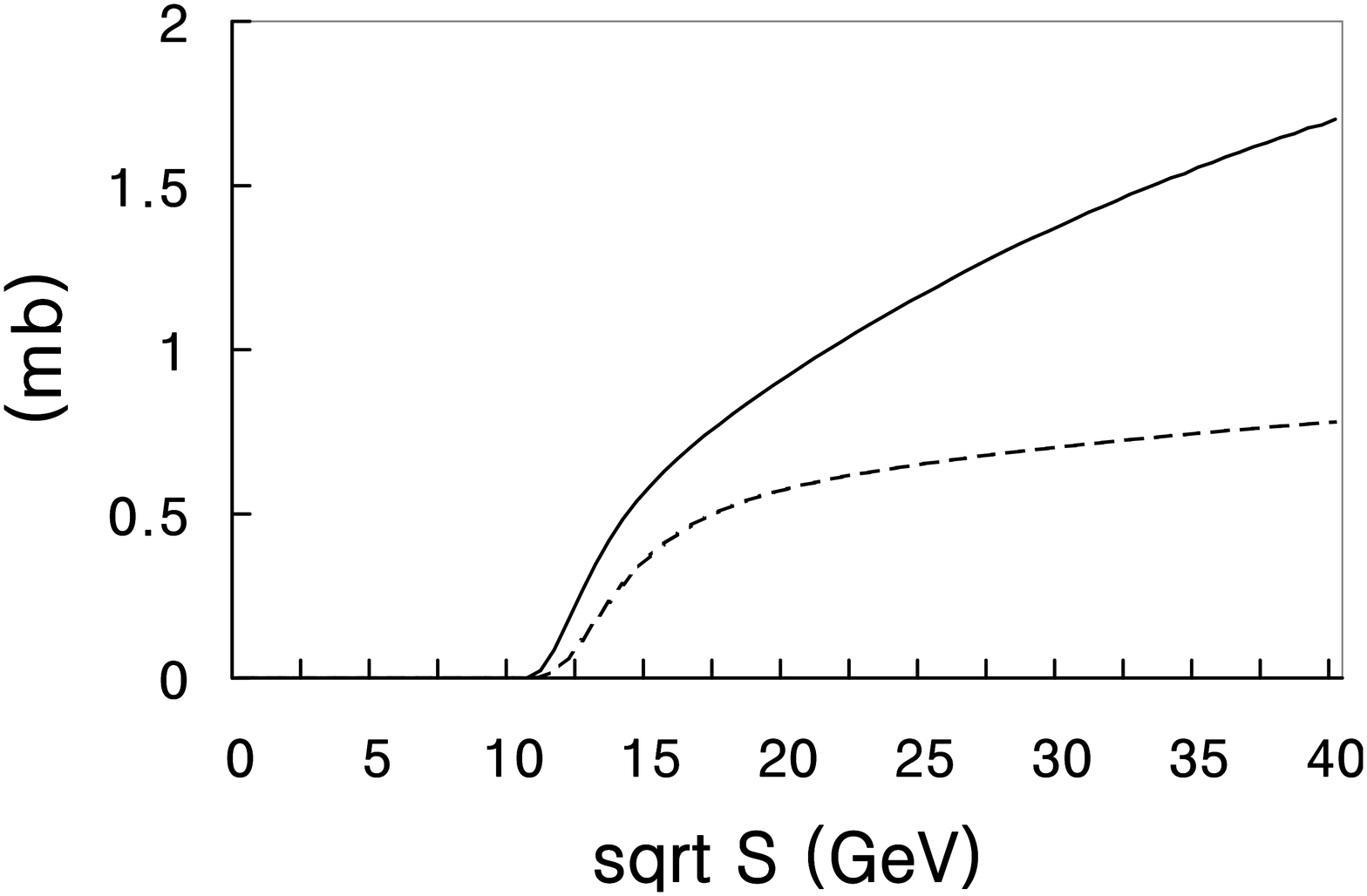}\hfill
\includegraphics[width=6cm, angle=270]{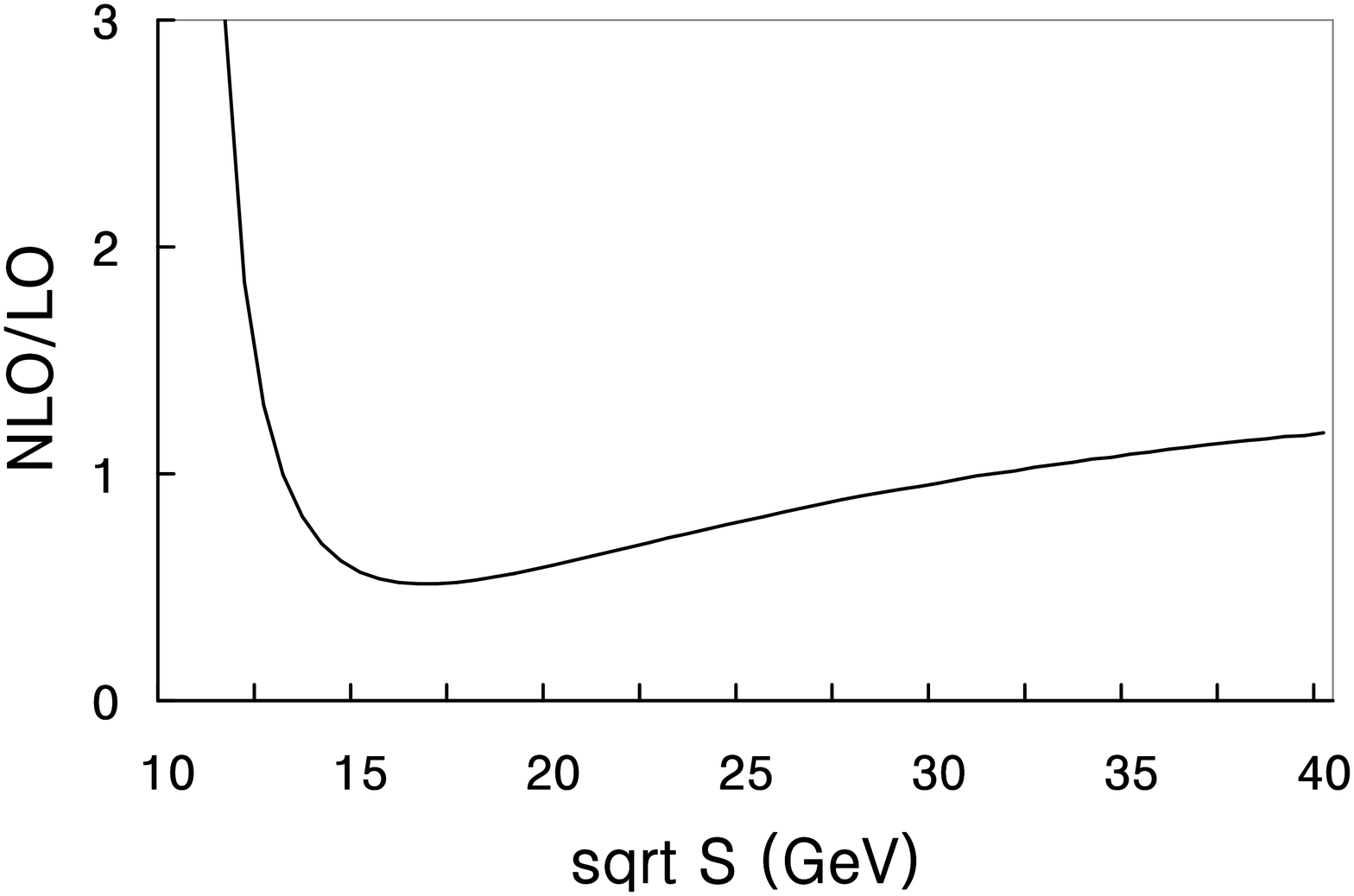}}\caption{The
left figure is the $\Upsilon$(1S)+nucleon total cross section to
LO (dashed line) and to  NLO (solid line).  The right figure is
the corresponding ratio between NLO and LO
results.}\label{fig:result}
\end{figure}

\begin{figure}
\centerline{
\includegraphics[width=6cm, angle=270]{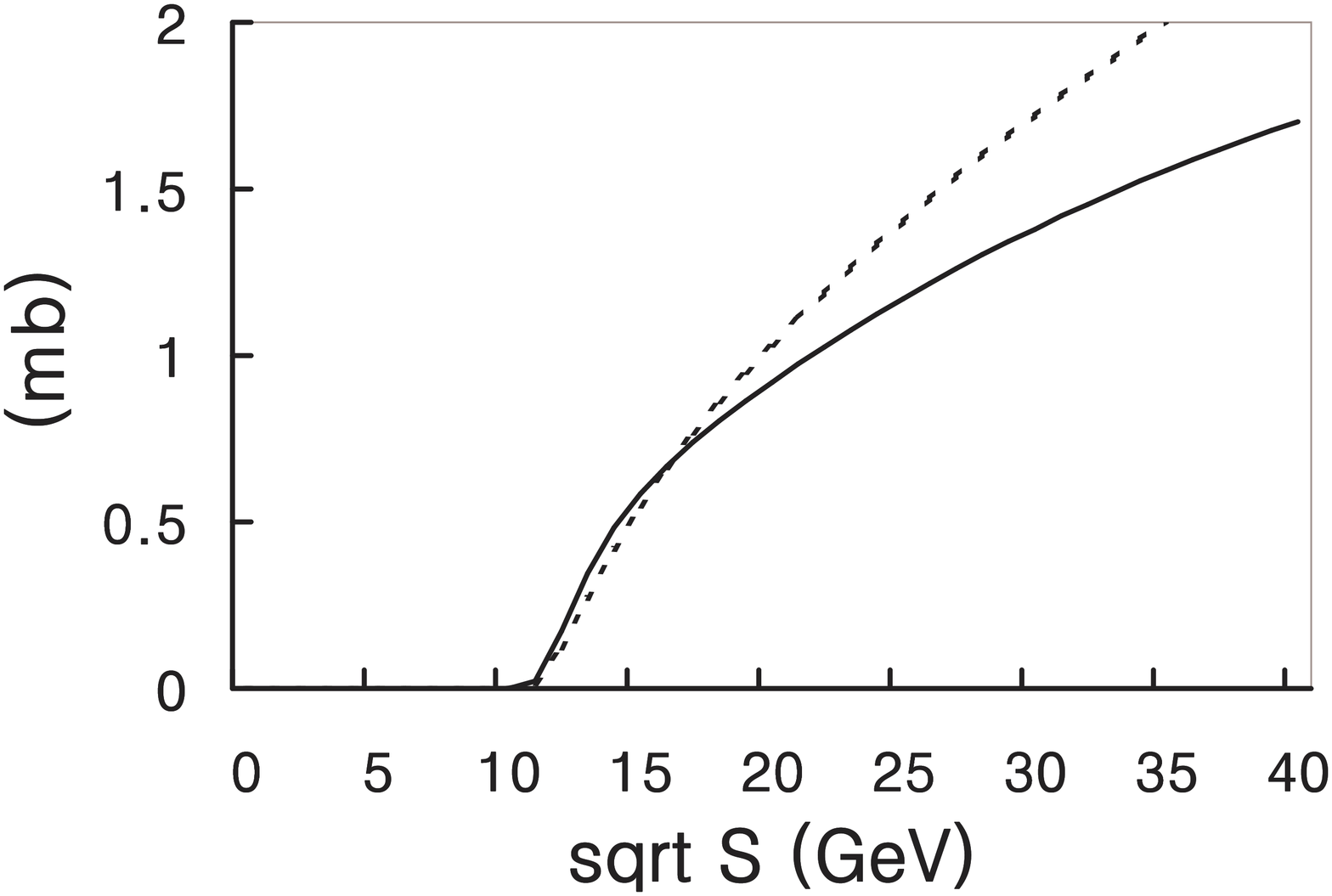}\hfill
\includegraphics[width=6cm, angle=270]{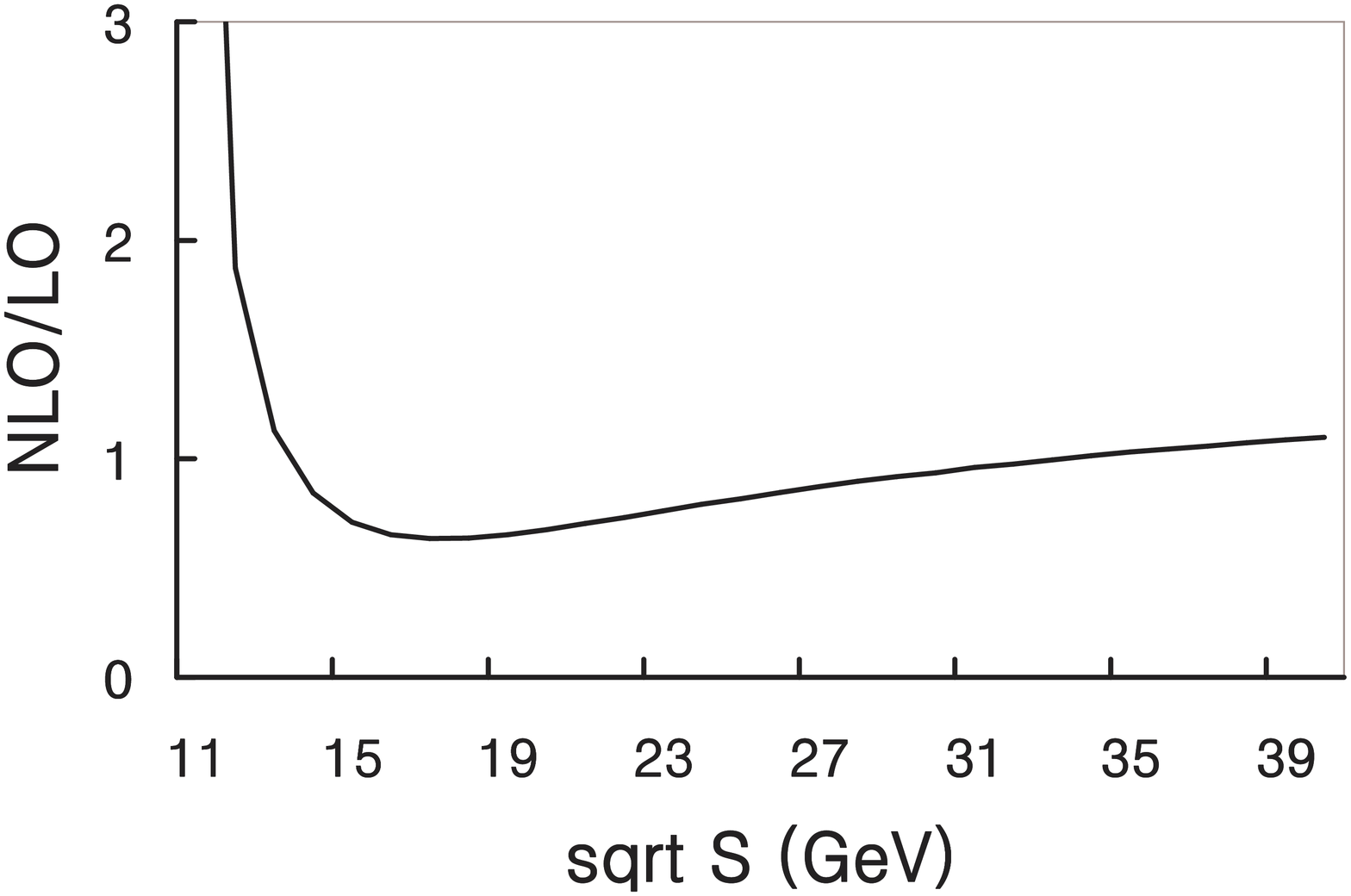}}\caption{The
left figure is the $\Upsilon$(1S)+nucleon total cross section with
$Q^2=1.25\ GeV^2$(straight line)
 and with $Q^2=2.0\ GeV^2$(dashed line).
 The right figure is the ratio between NLO and LO when
$Q^2=2.0 GeV^2$.}\label{fig:modi}
\end{figure}

We also applied the present calculation to the $J/\psi$
dissociation cross section. But in $J/\psi$ case, the soft plus
one loop correction has large negative value making the hadronic
cross section negative. This suggests that the formalism breaks
down for the charmonium system. As the quarkonium is heavier, the
relative contribution of this negative part is smaller. Therefore
we conclude that charm quark  is not heavy enough to use the
present formalism in the present form.

\section{Conclusion}

We have reported on the NLO calculation for the quarkonium parton
cross section in QCD.  All the collinear divergences have been
shown to cancel through mass factorization and the soft
divergences among themselves.   The result constitutes an exact
QCD calculation at the NLO in the formal heavy quark limit.

Explicit application to the Upsilon system shows that there are
large NLO corrections especially near the threshold, as has been
originally anticipated by Peskin\cite{peskin}.
Nevertheless, we have identified a window of energy range where
the NLO are under control, such that the perturbative QCD results
are reliable. Moreover, we have identified the origin of large
corrections and assessed the uncertainties through the magnitude
of the higher order correction.

The application to the charmonium system confirmed that the
discrepancies existing between LO QCD result and hadronic model
result on the charmonium dissociation cross section by hadrons
especially near threshold are partly due to large higher order
corrections in QCD.   Nevertheless, since the separation scale is
in the order of the binding energy, a thermal mass of few hundred
MeV for the partons will be enough to soften the NLO correction
and to make a perturbative treatment meaningful at finite
temperature.

\section{Acknowledgment}
Authors are grateful to W. Beenakker for his kind help, and also
to T. Hatsuda, and C. Y. Wong for useful discussion. This work was
supported by KOSEF under grant number M02-2004-000-10484-0

\appendix
\section{}

Here the phase space of 2-body and 3-body decay are reviewed. The
initial flux F is

\begin{eqnarray}
F=4\sqrt{(q\cdot k_1)^2-m_q^2 m_{k_1}^2}=2\acute{s},
\end{eqnarray}

and the total cross section of 2-body decay is

\begin{eqnarray}
\sigma&=&\frac{\mu^{-D+4}}{2\acute{s}}\int \frac{d^D
p_1}{(2\pi)^{D-1}}\int \frac{d^D
p_2}{(2\pi)^{D-1}}\delta^+(p_1^2-m^2)
\delta^+(p_2^2-m^2)\nonumber\\
&&\times (2\pi)^D \delta^D
(p_1+p_2-q-k_1)\overline{|M|^2}\nonumber\\
&=&\frac{\mu^{-D+4}}{2\acute{s}}\int \frac{d^D p_2}{(2\pi)^{D-2}}
\delta^+(p_2^2-m^2) \delta^+((q+k_1-p_2)^2-m^2)\overline{|M|^2}.
\end{eqnarray}
Because

\begin{eqnarray}
\int d^D p_2 \delta^+ (p_2^2-m^2)=\frac{1}{2}\int dp_{20}
d|\vec{p_2}|^2 |\vec{p_2}|^{D-3} d\Omega_{D-2}\delta^+
(p_{20}^2-|\vec{p_2}|^2-m^2)\nonumber\\
=\frac{\pi^{D/2-1}}{\Gamma(D/2-1)}\int dE_2 d\cos \chi
(E_2^2-m^2)^{\frac{D-3}{2}}(1-\cos^2 \chi)^{\frac{D-4}{2}},
\end{eqnarray}
where we have used the formula

\begin{eqnarray}
\int^\pi_0 \sin^D
d\theta=\sqrt{\pi}\frac{\Gamma(D/2+1/2)}{\Gamma(D/2+1)},
\end{eqnarray}
and

\begin{eqnarray}
d\Omega_{D-2}&=&\sin^{D-3}\theta_1 \sin^{D-4}\theta_2 ... \sin
\theta_{D-3} d\theta_1 d\theta_2 ... d\theta_{D-2}\nonumber\\
&=&\frac{2\pi^{D/2-1}}{\Gamma(D/2-1)}\sin^{D-3} \theta_1
d\theta_1,
\end{eqnarray}
the cross section becomes

\begin{eqnarray}
\sigma
&=&\frac{\mu^{-D+4}}{2\acute{s}}\frac{1}{(2\pi)^{D-2}}\frac{\pi^{D/2-1}}{\Gamma(D/2-1)}\int
dE_2 d\cos \chi (E_2^2-m^2)^{\frac{D-3}{2}}(1-\cos^2
\chi)^{\frac{D-4}{2}}\nonumber\\
&&\ \ \ \ \ \ \ \ \ \ \ \ \ \ \ \ \ \ \ \ \ \ \ \ \ \ \ \ \ \ \ \
\ \ \ \ \ \ \ \times \delta (s+t_1+u_1)\overline{|M|^2},
\end{eqnarray}
where $(q+k_1-p_2)^2-m^2=s+t_1+u_1$.

In the center of mass frame of q and $k_1$,

\begin{eqnarray}
q&=&(E_q, 0, ... , 0, k_{10})\nonumber\\
k_1&=&(k_{10}, 0, ... , 0, -k_{10})\nonumber\\
p_2&=&(E_2, 0, ... , |\vec{p}|\sin \chi, |\vec{p}|\cos \chi),
\end{eqnarray}
and
\begin{eqnarray}
E_q&=&\frac{s+m_\Phi^2}{2\sqrt{s}}\nonumber\\
k_{10}&=&\frac{\acute{s}}{2\sqrt{s}}\nonumber\\
E_2&=&-\frac{t_1+u_1}{2\sqrt{s}}\nonumber\\
|\vec{p}|&=&\frac{\sqrt{(t_1+u_1)^2-4sm^2}}{2\sqrt{s}}\nonumber\\
\cos \chi &=& \frac{2u_1 s
-(s+m_\Phi^2)(t_1+u_1)}{\acute{s}\sqrt{(t_1+u_1)^2-4sm^2}}.
\label{angle}
\end{eqnarray}

From the above relations, the Jacobian from $u_1,t_1$ to $E_2,
\cos \chi$ is

\begin{eqnarray}
dE_2 d\cos \chi =
\frac{\sqrt{s}}{\acute{s}\sqrt{(t_1+u_1)^2-4sm^2}}dt_1 du_1.
\end{eqnarray}

Then the differential cross section is

\begin{eqnarray}
\acute{s}^2\frac{d^2 \sigma}{dt_1
du_1}=\frac{\pi}{\Gamma(D/2-1)}\bigg(\frac{1}{4\pi}\bigg)^{D/2}
\bigg[\frac{\acute{s}u_1 t_1-m_\Phi^2 t_1^2 -m^2
\acute{s}^2}{\mu^2
\acute{s}^2}\bigg]^{\frac{D-4}{2}}\nonumber\\
\times \delta(s+t_1+u_1) \overline{|M|^2}. \label{2-body}
\end{eqnarray}

Next, the 3-body phase space is

\begin{eqnarray}
\sigma &=& \frac{\mu^{-2(D-4)}}{2\acute{s}}\int \frac{d^D
p_2}{(2\pi)^{D-1}}\int \frac{d^D p_1}{(2\pi)^{D-1}}\int \frac{d^D
k_2}{(2\pi)^{D-1}}\delta^+ (p_2^2-m^2)\nonumber\\ &&\times
\delta^+ (p_1^2-m^2)\delta^+ (k_2^2) (2\pi)^D \delta^D
(q+k_1-p_1-p_2-k_2) \bar{|M|}^2\nonumber\\
\nonumber\\
&=& \frac{1}{2\acute{s}} \frac{\mu^{-2(D-4)}}{(2\pi)^{2D-3}}\int
d^D
p_2 d^D p \delta^+ (p_2^2-m^2) \delta^D (q+k_1-p_2-p)\nonumber\\
&&\times \int d^D k_2 d^D p_1 \delta^+ (p_1^2-m^2) \delta^+
(k_2^2) \delta^D (p-k_2-p_1)\bar{|M|}^2.
\end{eqnarray}

In the above equation, new variable p is introduced and
D-dimensional p integration and D-dimension delta function of p
are inserted into phase space. The last line is the same as the
two body phase space, and it becomes

\begin{eqnarray}
\frac{-\pi^{\frac{D}{2}}}{2}\frac{\Gamma(\frac{D}{2}-1)}{\Gamma(D-3)}
\frac{(p^2-m^2)^{D-3}}{(p^2)^{\frac{D}{2}-1}}\int^\pi_0 d\theta_1
\sin^{D-3} \theta_1 \int^\pi_0 \sin^{D-4} \theta_2 \bar{|M|}^2.
\end{eqnarray}

Moving to $q$ and $k_1$ center of mass frame, the differential
cross section for three body decay is

\begin{eqnarray}
\acute{s}^2 \frac{d^2 \sigma}{dt_1 du_1} &=&
\frac{1}{2}\frac{1}{(4\pi)^D}\frac{\mu^{-D+4}}{\Gamma(D-3)}\bigg(\frac{\acute{s}u_1
t_1-m_\Phi^2 t_1^2-m^2\acute{s}^2}{\acute{s}^2 \mu^2 }
\bigg)^{\frac{D-4}{2}}
\frac{s_4^{D-3}}{(s_4+m^2)^{\frac{D}{2}-1}}\nonumber\\
&&\times \int^\pi_0 d\theta_1 \sin^{D-3}\theta_1 \int^\pi_0
d\theta_2 \sin^{D-4}\theta_2 \overline{|M|}^2.
\end{eqnarray}

\section{}
Here the angular integration which are used throughout this work
is derived. The required angular integration has the form

\begin{eqnarray}
I_D^{(i,j)}&\equiv& \int^\pi_0 d\theta_1 \sin^{D-3}\theta_1
\int^\pi_0 d\theta_2 \sin^{D-4}\theta_2\frac{1}{\acute{t}^i(\acute{u}+\acute{s})^j}\nonumber\\
&=& \int^\pi_0 d\theta_1  \int^\pi_0 d\theta_2
\frac{\sin^{D-3}\theta_1 \sin^{D-4}\theta_2
}{(a+b\cos\theta_1)^i(A+B\cos\theta_1+C\sin\theta_1
\cos\theta_2)^j},
\end{eqnarray}
where

\begin{eqnarray}
A &=&
2E_q(k_{10}-k_{20})+2k_{10}(k_{10}-|\vec{p}|\cos\varphi)\nonumber\\
B &=& 2k_{20}(|\vec{p}|\cos\varphi-k_{10})\nonumber\\
C &=& 2k_{20}|\vec{p}|\sin \varphi \nonumber\\
a &=& -2k_{10}k_{20}.
\end{eqnarray}
D denote that it is D-dimensional angular integration. If $b=-a,
A^2=B^2+C^2$, the solution has an explicit form
\cite{Beenakker}\cite{neerven}.

\begin{eqnarray}
I_D^{(i,j)}=\frac{2\pi}{(2a)^i
(2A)^j}\frac{\Gamma(D/2-i-1)\Gamma(D/2-j-1)\Gamma(D-3)}{\Gamma^2(D/2-1)\Gamma(D-i-j-2)}
\nonumber\\
\times F_{1,2}\bigg[
i,j,\frac{D}{2}-1;\frac{A-B}{2A}\bigg].\label{simplest}
\end{eqnarray}

$F_{1,2}$ is the hypergeometric function.

\begin{eqnarray}
F_{1,2}(a,b,c;x)\equiv
\frac{\Gamma(c)}{\Gamma(b)\Gamma(c-b)}\int^1_0 dt\
t^{b-1}(1-t)^{c-b-1}(1-tx)^{-a},
\end{eqnarray}
which has the following properties

\begin{eqnarray}
F_{1,2}(a,b,c;x)&=&F_{1,2}(b,a,c;x)\nonumber\\
F_{1,2}(0,b,c;x)&=&1.
\end{eqnarray}

In the present case $b=-a$, and $A^2 \neq B^2+C^2$. First,
considering $I_D^{(1,1)}$,

\begin{eqnarray}
I_D^{(1,1)}&=&
\int^\pi_0 d\theta_1 \int^\pi_0 d\theta_2 \frac{\sin^{D-3}\theta_1
\sin^{D-4}\theta_2}{a(1-\cos\theta_1)(A+B\cos\theta_1+C\sin\theta_1
\cos\theta_2)}\nonumber\\
&=&\frac{1}{a(A+B)}\int^\pi_0 d\theta_1 \int^\pi_0
d\theta_2  \frac{\sin^{D-3}\theta_1 \sin^{D-4}\theta_2}{(1-\cos\theta_1)}\nonumber\\
&&-\frac{1}{a(A+B)}\int^\pi_0 d\theta_1 \int^\pi_0 d\theta_2
\frac{\sin^{D-3}\theta_1
\sin^{D-4}\theta_2}{(1-\cos\theta_1)}\nonumber\\
&&\ \ \ \ \ \ \ \ \ \ \ \ \ \times \bigg[
1-\frac{A+B}{A+B\cos\theta_1+C\sin\theta_1
\cos\theta_2}\bigg]\nonumber\\
&\equiv& I_1 +I_2,
\end{eqnarray}
where $I$ is separated into infinite part $I_1$ and finite part
$I_2$ in $4$ dimension.

From Eq.~(\ref{simplest})

\begin{eqnarray}
I_1=\frac{I_D^{(1,0)}}{A+B}=\frac{\pi}{a(A+B)}\frac{2}{D-4},
\end{eqnarray}
and

\begin{eqnarray}
I_2&\approx& -\frac{1}{a(A+B)} \int^1_{-1} d\cos \theta_1
\int^\pi_0
d\theta_2 \frac{1}{(1-\cos\theta_1)}\nonumber\\
&&\times \bigg[ 1-\frac{A+B}{A+B\cos\theta_1+C\sin\theta_1
\cos\theta_2}\bigg] +\Theta(D-4)\nonumber\\
&=&-\frac{\pi}{a(A+B)} \int^1_{-1}\frac{d\cos
\theta_1}{1-\cos\theta_1} \nonumber\\
&&\times \bigg[
1-\frac{A+B}{\sqrt{(A+B\cos\theta_1)^2-C^2\sin^2\theta_1
}}\bigg] +\Theta(D-4)\nonumber\\
&=&\frac{\pi}{a(A+B)} \ln \bigg[ (A+B)^2-(1-\cos
\theta_1)(B^2+C^2+AB)\nonumber\\
&&+(A+B)\sqrt{(A+B\cos\theta_1)^2+C^2\sin^2 \theta_1}\bigg]
\bigg|^{\cos\theta_1=+1}_{\cos\theta_1=-1}+\Theta(D-4)\nonumber\\
&=&\frac{\pi}{a(A+B)} \ln\frac{(A+B)^2}{A^2-B^2-C^2}+\Theta(D-4).
\end{eqnarray}

As a result,

\begin{eqnarray}
I_D^{(1,1)}=\frac{\pi}{a(A+B)}\bigg[\frac{2}{D-4}+\ln\frac{(A+B)^2}{A^2-B^2-C^2}
\bigg]+\Theta(D-4).
\end{eqnarray}

$I_D^{(1,j)}$ with higher order $j$ is derived by differentiating
$I_D^{(1,1)}$ with respect to $A$.

\begin{eqnarray}
I_D^{(1,j)}=\frac{(-1)^{j-1}}{(j-1)!}\frac{d^{j-1}}{dA^{j-1}}I_D^{(1,1)}.
\end{eqnarray}

Below are the summary of the results. (the subscript `D' is
omitted for simplicity).

\begin{eqnarray}
I^{(1,0)}=\frac{2\pi}{a}\frac{1}{D-4}
\end{eqnarray}

\begin{eqnarray}
I^{(1,1)}&=&\frac{\pi}{a(A+B)}\bigg(\frac{2}{D-4}+\ln\bigg[\frac{(A+B)^2}{A^2-B^2-C^2}\bigg]\nonumber\\
&&+\frac{D-4}{2}\bigg[\ln^2
\frac{A-\sqrt{B^2+C^2}}{A+B}-\frac{1}{2}\ln^2\frac{A+\sqrt{B^2+C^2}}{A-\sqrt{B^2+C^2}}\nonumber\\
&&+2Li_2\bigg(-\frac{B+\sqrt{B^2+C^2}}{A-\sqrt{B^2+C^2}}\bigg)
-2Li_2 \bigg(\frac{A-\sqrt{B^2+C^2}}{A+B}\bigg)\bigg]\nonumber\\
&&+O((D-4)^2)\bigg)
\end{eqnarray}

\begin{eqnarray}
I^{(1,2)}=\frac{\pi}{a(A+B)^2}\bigg(\frac{2}{D-4}+\ln\bigg[\frac{(A+B)^2}{A^2-B^2-C^2}\bigg]
+O(D-4)\bigg)
\end{eqnarray}

\begin{eqnarray}
I^{(1,3)}=\frac{\pi}{a(A+B)^3}\bigg(
\frac{2}{D-4}+\ln\bigg[\frac{(A+B)^2}{A^2-B^2-C^2}\bigg]+\frac{2A(A+B)}{A^2-B^2-C^2}\nonumber\\
+(A+B)^2\frac{A^2+B^2+C^2}{(A^2-B^2-C^2)^2}-3+O(D-4)\bigg)
\end{eqnarray}

\begin{eqnarray}
I^{(1,4)}&=&\frac{\pi}{a(A+B)^4}\bigg(
\frac{2}{D-4}+\ln\bigg[\frac{(A+B)^2}{A^2-B^2-C^2}\bigg]+\frac{2A(A+B)}{A^2-B^2-C^2}\nonumber\\
&&+(A+B)^2\frac{A^2+B^2+C^2}{(A^2-B^2-C^2)^2}
+(A+B)^3\frac{2A^3+6AB^2+6AC^2}{3(A^2-B^2-C^2)^3}\nonumber\\
&&-11/3+O(D-4)\bigg)
\end{eqnarray}

\begin{eqnarray}
I^{(1,5)}&=&\frac{\pi}{a(A+B)^5}\bigg(
\frac{2}{D-4}+\ln\bigg[\frac{(A+B)^2}{A^2-B^2-C^2}\bigg]+\frac{2A(A+B)}{A^2-B^2-C^2}\nonumber\\
&&+(A+B)^2\frac{A^2+B^2+C^2}{(A^2-B^2-C^2)^2}
+(A+B)^3\frac{2A(A^2+3B^2+3C^2)}{3(A^2-B^2-C^2)^3}\nonumber\\&&+(A+B)^4
\bigg[\frac{4A^2(B^2+C^2)}{(A^2-B^2-C^2)^4}+\frac{1}{(A^2-B^2-C^2)^2}\bigg]\nonumber\\
&&-25/4+O(D-4)\bigg)
\end{eqnarray}

\begin{eqnarray}
 I^{(1,6)}&=&\frac{\pi}{a(A+B)^6}\bigg(
\frac{2}{D-4}+\ln\bigg[\frac{(A+B)^2}{A^2-B^2-C^2}\bigg]+\frac{2A(A+B)}{A^2-B^2-C^2}\nonumber\\
&&+(A+B)^2\frac{A^2+B^2+C^2}{(A^2-B^2-C^2)^2}
+(A+B)^3\frac{2A(A^2+3B^2+3C^2)}{3(A^2-B^2-C^2)^3}\nonumber\\&&+(A+B)^4
\bigg[\frac{4A^2(B^2+C^2)}{(A^2-B^2-C^2)^4}+\frac{1}{(A^2-B^2-C^2)^2}\bigg]\nonumber\\
&&+(A+B)^5\bigg[\frac{32A^5}{5(A^2-B^2-C^2)^5}-\frac{2A(3A^2+B^2+C^2)}{(A^2-B^2-C^2)^4}\bigg]\nonumber\\
&&-137/30+O(D-4)\bigg)
\end{eqnarray}

$I^{(0,j)}$ and $I^{(-1,j)}$ may be obtained from \cite{Beenakker}
and by the same method,

\begin{eqnarray}
I^{(0,1)}=\frac{\pi\ln\bigg(\frac{A+\sqrt{B^2+C^2}}{A-\sqrt{B^2+C^2}}\bigg)}{\sqrt{B^2+C^2}}
\end{eqnarray}

\begin{eqnarray}
I^{(0,2)}=\frac{2\pi}{A^2-B^2-C^2}
\end{eqnarray}

\begin{eqnarray}
I^{(0,3)}=\frac{2\pi A}{(A^2-B^2-C^2)^2}
\end{eqnarray}

\begin{eqnarray}
I^{(0,4)}=-\frac{2\pi(3A^2+B^2+C^2)}{3(A^2-B^2-C^2)^3}
\end{eqnarray}

\begin{eqnarray}
I^{(0,5)}=\frac{2\pi A(A^2+B^2+C^2)}{(A^2-B^2-C^2)^4}
\end{eqnarray}

\begin{eqnarray}
I^{(0,6)}=\frac{2\pi(5A^2+10A^2(B^2+C^2)+(B^2+C^2)^2)}{5(A^2-B^2-C^2)^5}
\end{eqnarray}

\begin{eqnarray}
I^{(-1,1)}=\pi a \bigg(
-\frac{2B}{B^2+C^2}+\frac{B^2+C^2+AB}{(B^2+C^2)^{3/2}}\ln
\bigg[\frac{A+\sqrt{B^2+C^2}}{A-\sqrt{B^2+C^2}}\bigg]\bigg)
\end{eqnarray}

\begin{eqnarray}
I^{(-1,2)}&=&\pi a \bigg(
\frac{2(AB+B^2+C^2)}{(B^2+C^2)(A^2-B^2-C^2)}\nonumber\\
&&-\frac{B}{(B^2+C^2)^{3/2}}\ln
\bigg[\frac{A+\sqrt{B^2+C^2}}{A-\sqrt{B^2+C^2}}\bigg]\bigg)
\end{eqnarray}

\begin{eqnarray}
I^{(-1,3)}=\frac{2\pi a(A+B)}{(A^2-B^2-C^2)^2}
\end{eqnarray}

\begin{eqnarray}
I^{(-1,4)}=\frac{2\pi a (3A^2+4AB+B^2+C^2)}{3(A^2-B^2-C^2)^3}
\end{eqnarray}

\begin{eqnarray}
I^{(-1,5)}=\frac{2\pi
a(3A^3+5A^2B+(3A+B)(B^2+C^2))}{3(A^2-B^2-C^2)^4}
\end{eqnarray}

\begin{eqnarray}
&&I^{(-1,6)}=\nonumber\\
\nonumber\\
&&\frac{2\pi
a(5A^4+10A^3B+(10A^2+6AB)(B^2+C^2)+(B^2+C^2)^2)}{5(A^2-B^2-C^2)^5}.
\end{eqnarray}

\section{}
Here the detail calculation of Fig.~(\ref{fig:nlog}) is presented.
$M_1$ through $M_8$ correspond to tree diagrams (1) to (8).

$M_1$ is

\begin{eqnarray}
M_{\mu\nu\lambda}^{(1)}&=& (ig)^2
\bar{u}(p_1)\Gamma_\mu(p_1,p_1-q)i\Delta(p_1-q) \gamma_\nu T^a
\nonumber\\ &&\ \ \ \ \ \ \times i\Delta(-p_2-k_2)\gamma_\lambda
T^b
v(p_2)\nonumber\\
&=& g^2
\sqrt{\frac{m_\Phi}{N_c}}\bar{u}(p_1)\frac{1+\gamma^0}{2}\gamma_i
g^{i\mu}\frac{1-\gamma^0}{2}T^a T^b v(p_2)\nonumber\\
&&\times \psi(p)\bigg[ g^0_\nu g^0_\lambda
\frac{-1}{k_{20}}\nonumber\\
&&\ \ \ \ \ \ \ \ \ \ \ \ -g^0_\nu g^0_\lambda
\frac{\vec{p_2}\cdot\vec{k_2}}{mk_{20}^2}-g^0_\nu
\frac{p_{2j}g^j_\lambda}{mk_{20}}+g^0_\lambda
\frac{p_{1j}g^j_\nu}{mk_{20}}\bigg].
\end{eqnarray}

The first line in the bracket is of $1/(mg^4)$ order, and the
second line is of $1/(mg^2)$ order from Eq.~(\ref{gorder}). As
will be shown later, the sum of $1/(mg^4)$ order terms from $M_1$
to $M_8$ vanishes. As a result, the $1/(mg^2)$ order becomes the
leading order.

$M_2$ is

\begin{eqnarray}
M_{\mu\nu\lambda}^{(2)}&=&(ig)^2 \bar{u}(p_1)\gamma_\lambda T^b
i\Delta(p_1+k_2) \gamma_\nu T^a i\Delta(q-p_2)  \nonumber\\
&&\ \ \ \ \ \ \times \Gamma_\mu(-p_2,q-p_2) v(p_2)\nonumber\\
&=& g^2
\sqrt{\frac{m_\Phi}{N_c}}\bar{u}(p_1)\frac{1+\gamma^0}{2}\gamma_i
g^{i\mu}\frac{1-\gamma^0}{2}T^b T^a v(p_2)\nonumber\\
&&\times \bigg[ g_\nu^0 g_\lambda^0
\frac{-1}{k_{20}}\psi(p)\nonumber\\
&&\ \ \ \ \ \ \ \ \ \ \ \ +\bigg( g_\nu^j g_\lambda^0
\frac{p_{2j}}{mk_{20}}-g_\nu^0 g_\lambda^j
\frac{p_{1j}}{mk_{20}}-g_\nu^0 g_\lambda^0
\frac{\vec{p_1}\cdot\vec{k_2}}{mk_{20}^2}\bigg) \psi(p)
\nonumber\\
&&\ \ \ \ \ \ \ \ \ \ \ \ +g_\nu^0 g_\lambda^0
\frac{-1}{k_{20}}\frac{\partial \psi (p)}{\partial \vec{p}}\cdot
(-\vec{k_1}+\vec{k_2})\bigg].
\end{eqnarray}

The third line in the bracket comes from the expansion of the
Bethe-Salpeter amplitude.

\begin{eqnarray}
\Gamma_\mu (p_1+k_2,-p_2+k_1)&=&\Gamma_\mu (\vec{p_1}+\vec{k_2},
\vec{p_1}+\vec{k_2})\label{bsexpansion1}\nonumber\\
&=&\Gamma_\mu (\vec{p_1},\vec{p_1})+\frac{\partial \Gamma_\mu
(\vec{p_1},\vec{p_1})}{\partial \vec{p}}\cdot \vec{k_2}\\
\Gamma_\mu (-p_2,q-p_2) &=&\Gamma_\mu
(\vec{p_1}-\vec{k_1}+\vec{k_2},\vec{p_1}-\vec{k_1}+\vec{k_2})\label{bsexpansion2}\nonumber\\
&=&\Gamma_\mu (\vec{p_1},\vec{p_1})+\frac{\partial \Gamma_\mu
(\vec{p_1},\vec{p_1})}{\partial \vec{p}}\cdot (-\vec{k_1}+\vec{k_2})\\
\Gamma_\mu (p_1-k_1,-p_2-k_2)&=&\Gamma_\mu (\vec{p_1}-\vec{k_1},
\vec{p_1}-\vec{k_1})\nonumber\\
&=&\Gamma_\mu (\vec{p_1},\vec{p_1})+\frac{\partial \Gamma_\mu
(\vec{p_1},\vec{p_1})}{\partial \vec{p}}\cdot (-\vec{k_1}).
\end{eqnarray}

$M^3$ is

\begin{eqnarray}
M_{\mu\nu\lambda}^{(3)}&=&(ig)^2 \bar{u}(p_1)\gamma_\lambda T^b
i\Delta(p_1+k_2)\Gamma_\mu
(p_1+k_2,-p_2+k_1)\nonumber\\
&&\ \ \ \ \ \ \times i\Delta(-p_2+k_1)\gamma_\nu T^a v(p_2)\nonumber\\
&=& g^2
\sqrt{\frac{m_\Phi}{N_c}}\bar{u}(p_1)\frac{1+\gamma^0}{2}\gamma_i
g^{i\mu}\frac{1-\gamma^0}{2}T^b T^a v(p_2)\nonumber\\
&& \times \bigg[ g^0_\nu g^0_\lambda \frac{-\epsilon
+|\vec{p_1}|^2/m}{k_{10}k_{20}}\psi(p)\nonumber\\
&& \ \ \ \ \ \ +g^0_\nu g^0_\lambda \bigg(
-\frac{\vec{p_2}\cdot\vec{k_1}}{mk_{10}^2}+\frac{\vec{p_2}\cdot\vec{k_1}
-\vec{p_1}\cdot\vec{k_2}}{mk_{10}k_{20}}+\frac{\vec{p_1}\cdot\vec{k_2}}
{mk_{20}^2}\bigg)\psi(p)\nonumber\\
&&\ \ \ \ \ \ +\bigg(g^j_\nu g^0_\lambda
\frac{p_{2j}}{m}\frac{k_{10}-k_{20}}{k_{10}k_{20}} +g^0_\nu
g^j_\lambda
\frac{p_{1j}}{m}\frac{k_{10}-k_{20}}{k_{10}k_{20}}\bigg)\psi(p)\nonumber\\
&&\ \ \ \ \ \ +g^0_\nu g^0_\lambda \bigg(
\frac{2\vec{p_1}\cdot\vec{k_2}}{mk_{10}k_{20}}\psi(p)+\frac{k_{10}-k_{20}}{k_{10}k_{20}}
\frac{\partial \psi (p)}{\partial \vec{p}}\cdot
\vec{k_2}\bigg)\bigg] \nonumber\\
\nonumber\\
&=& g^2
\sqrt{\frac{m_\Phi}{N_c}}\bar{u}(p_1)\frac{1+\gamma^0}{2}\gamma_i
g^{i\mu}\frac{1-\gamma^0}{2}T^b T^a v(p_2)\nonumber\\
&& \times \bigg[ g^0_\nu g^0_\lambda \frac{k_{10}-k_{20}}{k_{10}k_{20}}\psi(p)\nonumber\\
&& \ \ \ \ \ \ +g^0_\nu g^0_\lambda \bigg(
-\frac{\vec{p_2}\cdot\vec{k_1}}{mk_{10}^2}+\frac{\vec{p_2}\cdot\vec{k_1}
-\vec{p_1}\cdot\vec{k_2}}{mk_{10}k_{20}}+\frac{\vec{p_1}\cdot\vec{k_2}}
{mk_{20}^2}\bigg)\psi(p)\nonumber\\
&&\ \ \ \ \ \ +\bigg(g^j_\nu g^0_\lambda
\frac{p_{2j}}{m}\frac{k_{10}-k_{20}}{k_{10}k_{20}} +g^0_\nu
g^j_\lambda
\frac{p_{1j}}{m}\frac{k_{10}-k_{20}}{k_{10}k_{20}}\bigg)\psi(p)\nonumber\\
&&\ \ \ \ \ \ +g^0_\nu g^0_\lambda \bigg(
\frac{\vec{p_1}\cdot\vec{k_1}+\vec{p_1}\cdot\vec{k_2}}{mk_{10}k_{20}}\psi(p)+\frac{k_{10}-k_{20}}{k_{10}k_{20}}
\frac{\partial \psi (p)}{\partial \vec{p}}\cdot
\vec{k_2}\bigg)\bigg],\nonumber\\
\end{eqnarray}

Notice that $|\vec{p_2}|^2\approx
|\vec{p_1}|^2-2\vec{p_1}\cdot(\vec{k_1}-\vec{k_2})$ from the order
counting of Eq.~(\ref{gorder}). The last line in the last equation
comes from the expansion of the Bethe-Salpeter amplitude in
Eq.~(\ref{bsexpansion1}).

Similarly, $M^4$ to $M^8$ are given as below.

\begin{eqnarray}
M_{\mu\nu\lambda}^{(4)}&=&(ig)^2 \bar{u}(p_1)\gamma_\nu T^a
i\Delta(p_1-k_1)\Gamma_\mu
(p_1-k_1,-p_2-k_2)\nonumber\\
&&\ \ \ \ \ \ \times i\Delta(-p_2-k_2)\gamma_\lambda T^b v(p_2)\nonumber\\
&=& g^2
\sqrt{\frac{m_\Phi}{N_c}}\bar{u}(p_1)\frac{1+\gamma^0}{2}\gamma_i
g^{i\mu}\frac{1-\gamma^0}{2}T^a T^b v(p_2)\nonumber\\
&& \times \bigg[ g^0_\nu g^0_\lambda \frac{-\epsilon
+|\vec{p_1}|^2/m}{k_{10}k_{20}}\psi(p)\nonumber\\
&& \ \ \ \ \ \ +g^0_\nu g^0_\lambda \bigg(
-\frac{\vec{p_1}\cdot\vec{k_1}}{mk_{10}^2}+\frac{\vec{p_1}\cdot\vec{k_1}
-\vec{p_2}\cdot\vec{k_2}}{mk_{10}k_{20}}+\frac{\vec{p_2}\cdot\vec{k_2}}
{mk_{20}^2}\bigg)\psi(p)\nonumber\\
&&\ \ \ \ \ \ +\bigg(g^j_\nu g^0_\lambda
\frac{p_{1j}}{m}\frac{k_{10}-k_{20}}{k_{10}k_{20}} +g^0_\nu
g^j_\lambda
\frac{p_{2j}}{m}\frac{k_{10}-k_{20}}{k_{10}k_{20}}\bigg)\psi(p)\nonumber\\
&&\ \ \ \ \ \ +g^0_\nu g^0_\lambda \bigg(
-\frac{2\vec{p_1}\cdot\vec{k_1}}{mk_{10}k_{20}}\psi(p)+\frac{k_{10}-k_{20}}{k_{10}k_{20}}
\frac{\partial \psi (p)}{\partial \vec{p}}\cdot
(-\vec{k_1})\bigg)\bigg] \nonumber\\
&=& g^2
\sqrt{\frac{m_\Phi}{N_c}}\bar{u}(p_1)\frac{1+\gamma^0}{2}\gamma_i
g^{i\mu}\frac{1-\gamma^0}{2}T^b T^a v(p_2)\nonumber\\
&& \times \bigg[ g^0_\nu g^0_\lambda \frac{k_{10}-k_{20}}{k_{10}k_{20}}\psi(p)\nonumber\\
&& \ \ \ \ \ \ +g^0_\nu g^0_\lambda \bigg(
-\frac{\vec{p_1}\cdot\vec{k_1}}{mk_{10}^2}+\frac{\vec{p_1}\cdot\vec{k_1}
-\vec{p_2}\cdot\vec{k_2}}{mk_{10}k_{20}}+\frac{\vec{p_2}\cdot\vec{k_2}}
{mk_{20}^2}\bigg)\psi(p)\nonumber\\
&&\ \ \ \ \ \ +\bigg(g^j_\nu g^0_\lambda
\frac{p_{1j}}{m}\frac{k_{10}-k_{20}}{k_{10}k_{20}} +g^0_\nu
g^j_\lambda
\frac{p_{2j}}{m}\frac{k_{10}-k_{20}}{k_{10}k_{20}}\bigg)\psi(p)\nonumber\\
&&\ \ \ \ \ \ -g^0_\nu g^0_\lambda \bigg(
\frac{\vec{p_1}\cdot\vec{k_2}+\vec{p_1}\cdot\vec{k_1}}{mk_{10}k_{20}}\psi(p)+\frac{k_{10}-k_{20}}{k_{10}k_{20}}
\frac{\partial \psi (p)}{\partial \vec{p}}\cdot
\vec{k_1}\bigg)\bigg]\nonumber\\
\end{eqnarray}

\begin{eqnarray}
M_{\mu\nu\lambda}^{(5)}&=& (ig)^2
\bar{u}(p_1)\Gamma_\mu(p_1,p_1-q)i\Delta(p_1-q) \gamma_\lambda
T^b\nonumber\\
&&\ \ \ \ \ \ \times i\Delta(-p_2+k_1)\gamma_\nu T^a v(p_2)\nonumber\\
&=& g^2
\sqrt{\frac{m_\Phi}{N_c}}\bar{u}(p_1)\frac{1+\gamma^0}{2}\gamma_i
g^{i\mu}\frac{1-\gamma^0}{2}T^b T^a v(p_2)\nonumber\\
&&\times \psi(p)\bigg[ g^0_\nu g^0_\lambda
\frac{1}{k_{10}}\nonumber\\
&&\ \ \ \ \ \ \ \ \ \ \ \ +g^0_\nu g^0_\lambda
\frac{\vec{p_2}\cdot\vec{k_1}}{mk_{10}^2}-g^0_\nu
\frac{p_{1j}g^j_\lambda}{mk_{10}}+g^0_\lambda
\frac{p_{2j}g^j_\nu}{mk_{10}}\bigg]
\end{eqnarray}

\begin{eqnarray}
M_{\mu\nu\lambda}^{(6)}&=&(ig)^2 \bar{u}(p_1)\gamma_\nu T^a
i\Delta(p_1-k_1) \gamma_\lambda T^b
 i\Delta(q-p_2)  \nonumber\\
&&\ \ \ \ \ \ \times \Gamma_\mu(q-p_2,-p_2) v(p_2)\nonumber\\
&=& g^2
\sqrt{\frac{m_\Phi}{N_c}}\bar{u}(p_1)\frac{1+\gamma^0}{2}\gamma_i
g^{i\mu}\frac{1-\gamma^0}{2}T^b T^a v(p_2)\nonumber\\
&&\times \bigg[ g_\nu^0 g_\lambda^0
\frac{1}{k_{10}}\psi(p)\nonumber\\
&&\ \ \ \ \ \ \ \ \ \ \ \ +\bigg( g_\nu^j g_\lambda^0
\frac{p_{1j}}{mk_{10}}-g_\nu^0 g_\lambda^j
\frac{p_{2j}}{mk_{10}}+g_\nu^0 g_\lambda^0
\frac{\vec{p_1}\cdot\vec{k_1}}{mk_{10}^2}\bigg) \psi(p)
\nonumber\\
&&\ \ \ \ \ \ \ \ \ \ \ \ +g_\nu^0 g_\lambda^0
\frac{1}{k_{10}}\frac{\partial \psi (p)}{\partial \vec{p}}\cdot
(-\vec{k_1}+\vec{k_2})\bigg]
\end{eqnarray}

\begin{eqnarray}
M_{\mu\nu\lambda}^{(7)}&=& ig
\bar{u}(p_1)\Gamma_\mu(p_1,p_1-q)i\Delta(p_1-q)\gamma^\delta T^c
v(p_2)\frac{-i}{(k_1-k_2)^2}
\nonumber\\
&&\times g f_{abc}\bigg( (k_1+k_2)_\delta g_{\nu
\lambda}+(-2k_2+k_1)_\nu g_{\lambda \delta}
+(-2k_1+k_2)_\lambda g_{\nu \delta}\bigg)  \nonumber\\
&=&
g^2\sqrt{\frac{m_\Phi}{N_c}}\bar{u}(p_1)\frac{1+\gamma^0}{2}\gamma_i
g^{i\mu}\frac{1-\gamma^0}{2}[T^a, T^b] v(p_2)\frac{1}{(k_1-k_2)^2}\nonumber\\
&&\times \bigg[ (k_1+k_2)_0 g_{\nu \lambda}+(-2k_2+k_1)_\nu
g_{\lambda 0} +(-2k_1+k_2)_\lambda g_{\nu 0}\nonumber\\
&&\ \ \ \ \ \
\frac{\vec{p_1}\cdot\vec{k_1}+\vec{p_1}\cdot\vec{k_2}}{m}g_{\nu
\lambda} +\frac{2k_{1j}p_{1k}+2k_{1k}p_{1j}}{m}g_\nu^j g_\lambda^k
\bigg]\psi(p)
\end{eqnarray}

\begin{eqnarray}
M_{\mu\nu\lambda}^{(8)}&=& ig \bar{u}(p_1)\gamma^\delta T^c
i\Delta(q-p_2) \Gamma_\mu(q-p_2,-p_2) v(p_2)\frac{-i}{(k_1-k_2)^2}
\nonumber\\
&&\times g f_{abc}\bigg( (k_1+k_2)_\delta g_{\nu
\lambda}+(-2k_2+k_1)_\nu g_{\lambda \delta}
+(-2k_1+k_2)_\lambda g_{\nu \delta}\bigg)  \nonumber\\
&=&
g^2\sqrt{\frac{m_\Phi}{N_c}}\bar{u}(p_1)\frac{1+\gamma^0}{2}\gamma_i
g^{i\mu}\frac{1-\gamma^0}{2}[T^a, T^b] v(p_2)\frac{1}{(k_1-k_2)^2}\nonumber\\
&&\times \bigg[ -\bigg((k_1+k_2)_0 g_{\nu \lambda}+(-2k_2+k_1)_\nu
g_{\lambda 0} +(-2k_1+k_2)_\lambda g_{\nu 0}\nonumber\\
&&\ \ \ \ \ \ \ \ \ \ \ \
\frac{\vec{p_2}\cdot\vec{k_1}+\vec{p_2}\cdot\vec{k_2}}{m}g_{\nu
\lambda} +\frac{2k_{1j}p_{2k}+2k_{1k}p_{2j}}{m}g_\nu^j g_\lambda^k
\bigg)\psi(p)\nonumber\\
&&\ \ \ \ \ \ +\bigg((k_1+k_2)_0 g_{\nu \lambda}+(-2k_2+k_1)_\nu
g_{\lambda 0} +(-2k_1+k_2)_\lambda g_{\nu 0}\bigg)\nonumber\\
&&\ \ \ \ \ \ \ \ \ \ \ \ \times \frac{\partial \psi(p)}{\partial
\vec{p}}\cdot (\vec{k_1}-\vec{k_2})\ \ \ \bigg]. \label{c11}
\end{eqnarray}

Next considering others,

\begin{eqnarray}
M_{\mu\nu\lambda}^{(9)}&=&(ig)^3 \int \frac{d^4
k}{(2\pi)^4}\gamma_\lambda T^b i\Delta(p_1+k_2)\gamma^\sigma T^c
i\Delta(k+q)\Gamma_\mu (k+q,k)\nonumber\\
&&\times i\Delta (k)\gamma^\delta T^d
v(p_2)\frac{-i}{(k+p_2)^2}\frac{-i}{(k+p_2-k_1)^2}\nonumber\\
&&\times g f^{cad}\bigg( (k+p_2-2k_1)_\delta g_{\sigma \nu}
+(k_1+k+p_2)_\sigma g_{\nu \delta} +(-2k-2p_2+k_1)_\nu g_{\sigma
\delta}\bigg)\nonumber\\
&=&
g^2\sqrt{\frac{m_\Phi}{N_c}}\bar{u}(p_1)\frac{1+\gamma^0}{2}\gamma_i
g^{i\mu}\frac{1-\gamma^0}{2}T^b T^a v(p_2)\nonumber\\
&&\times g_\nu^j g_\lambda^0 \frac{1}{k_{20}}\bigg[\
\frac{2p_{1j}}{m}\psi(p)-\bigg(\epsilon-\frac{|\vec{p}|^2}{m}
\bigg)\frac{\partial \psi(p)}{\partial p_j}\ \bigg]
\end{eqnarray}

\begin{eqnarray}
M_{\mu\nu\lambda}^{(10)}&=&(ig)^3 \int \frac{d^4
k}{(2\pi)^4}\gamma^\sigma T^c i\Delta (k)\Gamma_\mu
(k,k-q)i\Delta(k-q)\gamma^\delta T^d \nonumber\\
&&\times i\Delta(-p_2-k_2)\gamma_\lambda T^b v(p_2)
\frac{-i}{(k-p_1)^2}\frac{-i}{(k-p_1+k_1)^2}\nonumber\\
&&\times g f^{cad}\bigg( (k-p_1-k_1)_\delta g_{\sigma \nu}
+(2k_1+k-p_1)_\sigma g_{\nu \delta} +(-2k+2p_1-k_1)_\nu g_{\sigma
\delta}\bigg)\nonumber\\
&=&
g^2\sqrt{\frac{m_\Phi}{N_c}}\bar{u}(p_1)\frac{1+\gamma^0}{2}\gamma_i
g^{i\mu}\frac{1-\gamma^0}{2}T^a T^b v(p_2)\nonumber\\
&&\times g_\nu^j g_\lambda^0 \frac{1}{k_{20}}\bigg[\
-\frac{2p_{1j}}{m}\psi(p)+\bigg(\epsilon-\frac{|\vec{p}|^2}{m}
\bigg)\frac{\partial \psi(p)}{\partial p_j}\ \bigg]
\end{eqnarray}

\begin{eqnarray}
M_{\mu\nu\lambda}^{(11)}&=&(ig)^3 \int \frac{d^4
k}{(2\pi)^4}\gamma_\nu T^a i\Delta (p_1-k_1)\gamma^\sigma T^c
i\Delta(k+q) \Gamma_\mu
(k,k+q) \nonumber\\
&&\times i\Delta(k)\gamma^\delta T^d v(p_2)
\frac{-i}{(k+p_2)^2}\frac{-i}{(k+p_2+k_2)^2}\nonumber\\
&&\times g f^{cdb}\bigg( (2k+2p_2+k_2)_\lambda g_{\sigma \delta}
+(k_2-k-p_2)_\sigma g_{\lambda \delta} +(-k-p_2-2k_2)_\delta
g_{\sigma
\lambda}\bigg)\nonumber\\
&=&
g^2\sqrt{\frac{m_\Phi}{N_c}}\bar{u}(p_1)\frac{1+\gamma^0}{2}\gamma_i
g^{i\mu}\frac{1-\gamma^0}{2}T^a T^b v(p_2)\nonumber\\
&&\times g_\nu^0 g_\lambda^j \frac{1}{k_{10}}\bigg[\
-\frac{2p_{1j}}{m}\psi(p)+\bigg(\epsilon-\frac{|\vec{p}|^2}{m}
\bigg)\frac{\partial \psi(p)}{\partial p_j}\ \bigg]
\end{eqnarray}

\begin{eqnarray}
M_{\mu\nu\lambda}^{(12)}&=&(ig)^3 \int \frac{d^4
k}{(2\pi)^4}\gamma^\sigma T^c i\Delta(k)\Gamma_\mu (k,k-q)
i\Delta(k-q) \gamma^\delta T^d   \nonumber\\
&&\times i\Delta (-p_2+k_1)\gamma_\nu T^a v(p_2)
\frac{-i}{(k-p_1)^2}\frac{-i}{(k-p_1-k_2)^2}\nonumber\\
&&\times g f^{cdb}\bigg( (2k-2p_1-k_2)_\lambda g_{\sigma \delta}
+(2k_2-k+p_1)_\sigma g_{\lambda \delta} +(-k+p_1-k_2)_\delta
g_{\sigma
\lambda}\bigg)\nonumber\\
&=&
g^2\sqrt{\frac{m_\Phi}{N_c}}\bar{u}(p_1)\frac{1+\gamma^0}{2}\gamma_i
g^{i\mu}\frac{1-\gamma^0}{2}T^b T^a v(p_2)\nonumber\\
&&\times g_\nu^0 g_\lambda^j \frac{1}{k_{10}}\bigg[\
\frac{2p_{1j}}{m}\psi(p)-\bigg(\epsilon-\frac{|\vec{p}|^2}{m}
\bigg)\frac{\partial \psi(p)}{\partial p_j}\ \bigg]
\end{eqnarray}

\begin{eqnarray}
M_{\mu\nu\lambda}^{(16)}&=&(ig)^2 \int \frac{d^4
k}{(2\pi)^4}\gamma^\sigma T^c i\Delta(k)\Gamma_\mu (k,k-q)
i\Delta(k-q) \gamma^\delta T^d v(p_2)  \nonumber\\
&&\times
\frac{-i}{(k-p_1)^2}\frac{-i}{(k-p_1+k_1-k_2)^2}\frac{-i}{(k_1-k_2)^2}\nonumber\\
&&\times g f^{ced}\bigg( (k-p_1-k_1+k_2)_\delta g_{\sigma
\varepsilon} +(2k_1-2k_2+k-p_1)_\sigma g_{\varepsilon
\delta}\nonumber\\
&&\ \ \ \ \ \ \ \ \ \ \ \  +(-2k+2p_1-k_1+k_2)_\varepsilon
g_{\sigma
\delta}\bigg)\nonumber\\
&& \times g f^{eab}\bigg( (-2k_1+k_2)_\lambda g_{\nu \varepsilon}
+(k_1+k_2)_\varepsilon g_{\nu \lambda}+(k_1-2k_2)_\nu
g_{\varepsilon \lambda}\bigg)\nonumber\\
 &=&
g^2\sqrt{\frac{m_\Phi}{N_c}}\bar{u}(p_1)\frac{1+\gamma^0}{2}\gamma_i
g^{i\mu}\frac{1-\gamma^0}{2}\ [T^a, T^b]\ v(p_2)\frac{1}{(k_1-k_2)^2}\nonumber\\
&&\times \bigg( (-2k_1+k_2)_\lambda g_{\nu j} +(k_1+k_2)_j g_{\nu
\lambda}+(k_1-2k_2)_\nu
g_{j \lambda}\bigg)\nonumber\\
&&\times \bigg[\
\frac{2p_1^j}{m}\psi(p)-\bigg(\epsilon-\frac{|\vec{p}|^2}{m}
\bigg)\frac{\partial \psi(p)}{\partial p^j}\ \bigg]
\end{eqnarray}

Because $M^{(13)}\sim M^{(15)}$ are higher order  in the order
counting of Eq.~(\ref{gorder}) compared to the other terms, they
were ignored. Summing all amplitudes from $M^{(1)}$ to $M^{(16)}$,
the total invariant amplitude is given as,

\begin{eqnarray}
M^{\mu \nu \lambda } &=& \bigg[ \bigg(\frac{ \partial \psi (p
)}{\partial \vec{p}} \cdot \vec{k_1} \bigg) \bigg(-g^\lambda_0
g^\nu_0 \frac{1}{k_{20}} + \frac{1}{k_1 \cdot k_2 } (g^\lambda_0
k^\nu_2
+g^\nu_0 k^\lambda_1 - g^{\nu \lambda } k_{20}) \bigg) \nonumber\\
&&+\bigg(\frac{ \partial \psi (p )}{\partial \vec{p}} \cdot
\vec{k_2} \bigg) \bigg(g^\lambda_0 g^\nu_0 \frac{1}{k_{10}} -
\frac{1}{k_1 \cdot k_2 } (g^\lambda_0 k^\nu_2 +g^\nu_0 k^\lambda_1
- g^{\nu \lambda } k_{10}) \bigg)
\nonumber\\
&&+(k_{10 } -k_{20 }) \frac{\psi(p) }{\partial p_{j}}
\bigg(-g^\lambda_j g^\nu_0 \frac{1}{k_{10 }} -g^\lambda_0 g^\nu_j
\frac{1}{k_{20}} +\frac{1}{k_1 \cdot k_2 } (g^\nu_j k^\lambda_1
+g^\lambda_j k^\nu_2 )\bigg) \bigg] \nonumber\\
&&\ \ \ \ \ \ \times g^2 \sqrt{\frac{m_\Phi}{N_c }}
\overline{u}(p_1 )\frac{1+\gamma_0 }{2} \gamma_i g^{\mu
i}\frac{1-\gamma_0}{2}\ [T^a ,T^b ]\ v(p_2 ).
\end{eqnarray}

\section{}

In this appendix, we summarize the present counting scheme and
give details on determining the scaling properties of certain
diagrams.  To begin with, the binding energy of the quarkonium
scales as,
\begin{eqnarray}
\epsilon=m(N_cg^2/16\pi)^2 \sim O(mg^4).
\end{eqnarray}
In the quarkonium rest frame, the  energy conservation condition
for the process $\Phi+q,g(k_1) \rightarrow
Q(p_1)+\bar{Q(p_2)}+q,g(k_2)$ is,

\begin{eqnarray}
\epsilon_0=k_{10}-k_{20}-\frac{|\vec{p_1}|^2}{2m}-\frac{|\vec{p_2}|^2}{2m},
\end{eqnarray}
where $\epsilon_0$ is the binding energy of the quarkonium, and
$k_{10}$, $k_{20}$ are the energies of the incoming and outgoing
partons respectively. $\vec{p_1}$ and $\vec{p_2}$ are respectively
the three momenta of the heavy quark and antiquark from the
quarkonium. From this relation, the following order counting can
be deduced.
\begin{eqnarray}
|\vec{p_1}|\sim |\vec{p_2}|\sim O(mg^2), \ \ \ \ \ k_{10}\sim
k_{20}\sim O(mg^4).
\end{eqnarray}

The counting for the internal gluon loop momentum $K$, which
connects the heavy quark and antiquark within the bound state, can
be deduced from Eq.~(\ref{hardg}).  Since the left hand side of
Eq.~(\ref{hardg}) is of $O(mg^4)$, $K$ must be of $O(mg^2)$. In
contrast, the order of gluon momenta appearing in the perturbative
one loop correction should be of $O(mg^4)$.  This is so because
the separation scale, which sets the cut off in the perturbative
diagrams, are set to the binding energy, which is of $O(mg^4)$.
Within the bound state loop, the internal energy and heavy quark
propagator can be counted as $O(mg^4)$.

From the above considerations, the order of each Feynman rules can
be deduced and the results are listed in table
(\ref{tab:Feynman}).  Bound gluon means that it is the internal
gluon, which produces the coulomb bound state, and whose momentum
$K$ scales as $O(mg^2)$.   There are two types of three gluon
vertex.  In the first one, the vertex combines two bound gluons
and one external gluon, while in the other, it combines three
external gluons.

The order of a diagram can be deduced from the above order
counting scheme.   For example, the left and the right diagrams of
the Bethe-Salpeter equation in Fig.(\ref{fig:BS}) can be shown to
be of the same order.  Compared to the left diagram, the right
diagram has an addition internal loop ($ (mg^4)\times (mg^2)^3$),
two heavy quark propagator ($ (mg^4)^{-2}$), a bound gluon
propagator ($ (mg^2)^{-2}$), and two coupling constant ($g^2$),
which altogether gives order 1, as the left diagram.

The suppression of diagrams (13), (14) and (15) to the other
diagrams in Fig.~(\ref{fig:nlog}) may also be explained. Comparing
diagrams (13) or (14) to (9), diagram (9) has additionally a heavy
quark propagator ($(mg^4)^{-1}$) and a quark gluon vertex ($g$),
while diagram (13) or (14) has additionally a bound gluon
propagator ($(mg^2)^{-2}$) and a three gluon vertex (two bound
gluon + one external gluon of ($O(mg^3)$).  Hence diagram (13) or
(14) is suppressed by $O(g^2)$ relative to diagram (9). Similarly,
diagram (15) does not have a three gluon vertex ($O(mg^3)$) nor a
heavy quark propagator ($(mg^4)^{-1}$) but an additional four
gluon vertex ($g^2$) compared to diagram (9), and hence is
relatively suppressed by a factor of $g^2$.

In certain cases, the simple counting scheme has to be implemented
with care.   As an example, consider comparing the order of the
first two diagrams to the third diagram in Fig.~(\ref{fig:lo}).
Using our naive counting scheme above, the third diagram can be
shown to be suppressed by $g^2$ compared to the first two
diagrams. However, in this case, the first two diagrams cancel to
leading order in the counting and combine to give the same order
as the third diagram.

\begin{table}[t]
\centering
\begin{tabular}{ccc}
\hline Feynman diagram & order & reference \\
\hline heavy quark (antiquark) propagator & $(mg^4)^{-1}$ &
Eq.~(\ref{qpropa}) \\
bound gluon propagator & $(mg^2)^{-2}$ & Eq.~(\ref{hardg}) \\
external gluon momentum & $mg^4$ & Eq.~(\ref{gorder}) \\
bound gluon momentum & $mg^2$ & \\
 three gluon vertex (two bound and one
external gluons) &
$mg^3$ &\\
three gluon vertex (three external gluons) & $mg^5$ &\\
\hline
\end{tabular}
\bigskip
\caption{Order counting for various Feynman diagrams}
\label{tab:Feynman}
\end{table}

\end{document}